\documentclass[preprintnumbers,superscriptaddress,showkeys,showpacs,byrevtex]{revtex4}
\usepackage{amsmath,amsfonts,amssymb,amscd,amsxtra,amsthm}
\usepackage{graphicx}
\usepackage{bm}
\usepackage{epstopdf}
\usepackage{multirow}
\begin{document}
\preprint{KIAS-P12014}
\preprint{CYCU-HEP-12-01}
\title{Fragmentation and quark distribution functions for the pion and kaon\\
with explicit flavor-SU(3)-symmetry breaking}
\author{Seung-il Nam}
\email[E-mail: ]{sinam@kias.re.kr}
\affiliation{School of Physics, Korea Institute for Advanced Study (KIAS), Seoul 130-722, Korea}
\author{Chung-Wen Kao}
\email[E-mail: ]{cwkao@cycu.edu.tw}
\affiliation{Department of Physics, Chung-Yuan Christian University (CYCU), Chung-Li 32023, Taiwan}
\date{\today}
\begin{abstract}
We investigate the unpolarized pion and kaon fragmentation functions, employing the nonlocal chiral-quark model,
which manifests the nonlocal interactions between the quarks and pseudoscalar mesons, considering the explicit
flavor-SU(3)-symmetry breaking in terms of the current-quark masses. Moreover, we study the quark-distribution functions,
derived from the fragmentation ones with the Drell-Yan-Levi relation. Numerical results are evaluated to higher $Q^2$ by the
DGLAP evolution and compared with the empirical data. The ratios between the relevant valance quark-distribution functions
are also discussed. It turns out that the present results are in relatively good agreement with available data and other theoretical estimations.
\end{abstract}
\maketitle
\section{Introduction}
To apply perturbative Quantum Chromodynamics (QCD) to study hadronic processes such as the deep-inelastic
electron-proton scattering or semi-inclusive hadron production process,
provided that QCD factorization theorem is applicable there, the cross section is able to be expressed as a
convolution of the two parts: the process-dependent perturbative QCD(pQCD) calculable short-distance parton cross section,
and the universal long-distance functions which can be extracted from experiments, but
cannot be calculated by pQCD because the strong interaction in the long distance is nonperturbative.

These long-distance functions are the fundamental nonperturbative ingredients for the analysis of the scattering processes involving hadrons.
The fragmentation functions and the quark-distribution functions are belonged to this category.
The fragmentation functions, $D^h_q(z)$ indicate the probability for a hadron fragmented from a quark with the momentum fraction $z$, where the subscripts $h$ and $q$ indicate the hadron and quark, respectively. It plays an important role in analyzing the semi-inclusive processes in the electron-positron scattering, deep-inelastic proton-proton scattering, and so on~\cite{Collins:1992kk,Mulders:1995dh,Boer:1997nt,Anselmino:1994tv,Anselmino:2008jk,Christova:2006qs,Anselmino:2007fs,
Bacchetta:2006tn,Efremov:2006qm,Collins:2005ie,Ji:2004wu}. The parton distribution functions for hadrons $(h)$, $f^h_{q}(x)$ stand for the distribution of the momentum fraction $x$ carried by a parton inside a hadron. In analyzing the deep-inelastic electron-proton scattering, the parton distribution functions becomes crucial for instance. We note that these two nonperturbative functions are inter-related analytically via the Drell-Levy-Yan (DLY) relation~\cite{Drell:1969jm}. Note that this analyticity is possible only if those functions can be described by the same function, defined in the different regions~\cite{Drell:1969jm}.

Those functions have been studied intensively for several decades but still not fully-understood~\cite{Kretzer:2000yf,Conway:1989fs,deFlorian:2007aj,Sutton:1991ay,Hirai:2007cx,Kniehl:2000fe,Ji:1993qx,Jakob:1997wg, Bacchetta:2002tk,Amrath:2005gv,Bacchetta:2007wc,Meissner:2010cc,Bentz:1999gx,Ito:2009zc,Matevosyan:2010hh,Matevosyan:2011ey,
Matevosyan:2011vj,Nguyen:2011jy,Aloisio:2003xj,Dorokhov:1991nj,Nam:2006sx,Nam:2006au,Praszalowicz:2001pi,Noguera:2005cc,
Aicher:2010cb,Bacchetta:2006un,Holt:2010vj}. Empirically, their information can be extracted from the available high-energy
lepton-scattering data by global analyses with appropriate parameterizations satisfying certain constraints~\cite{Kretzer:2000yf,Conway:1989fs,deFlorian:2007aj,Sutton:1991ay,Hirai:2007cx,Kniehl:2000fe}.

From theoretical points of view, there have been numerous works done for those functions so far: The momentum sum rules for the fragmentation functions were proven in terms of QCD in a rigorous manner~\cite{Meissner:2010cc}. In Refs.~\cite{Ito:2009zc,Matevosyan:2010hh,Matevosyan:2011ey,Matevosyan:2011vj}, the Nambu-Jona-Lasinio (NJL) model was applied for these functions with the quark-jets and resonances, which satisfied the momentum sum rules. It turned out that the quark-jet contributions provide considerable contributions to the various fragmentation functions at the small $z$ region. The Dyson-Schwinger (DS) method presented considerably successful results for the valance-quark distribution functions for the pion and kaon~\cite{Nguyen:2011jy}. Monte-Carlo simulations with supersymmetric (SUSY) QCD were carried out to obtain the fragmentation function up to a very high energy in the center-of-mass frame $\sqrt{s}$~\cite{Aloisio:2003xj}.
The collins fragmentation functions which play an important role in the transverse-spin physics have also been studied in the quark-PS meson coupling model~\cite{Bacchetta:2002tk,Amrath:2005gv,Bacchetta:2007wc}. Dihadron fragmentation functions were also studied in the same theoretical formalism~\cite{Bacchetta:2006un}.

Similarly, the quark-distribution function has also been studied with a nonlocal effective-chiral Lagrangian, employing the DS method~\cite{Noguera:2005cc}. In Ref.~\cite{Aicher:2010cb}, the authors made use of the the Drell-Yan process including the soft-gluon resumption for the valance-quark distributions. A review for the experimental and theoretical status for valance-quark distribution for the nucleon and pion is given in Ref.~\cite{Holt:2010vj}.

In Refs.~\cite{Dorokhov:1991nj,Nam:2006sx,Nam:2006au,Praszalowicz:2001pi}, the instanton-motivated approaches were taken into account for computing the quark distribution functions, manifesting the nonlocal quark-pseudoscalar (PS) meson interactions. Employing the nonlocal chiral quark model
(NLChQM) which is motivated by the dilute instanton-liquid model (LIM), which is properly defined in Euclidean space in principle~\cite{Shuryak:1981ff,Diakonov:1985eg,Diakonov:1983hh,Schafer:1996wv,Diakonov:2002fq}. NLChQM have accumulated successful results for various nonperturbative quantities in good agreement with experiments as well as lattice QCD (LQCD) simulations.~\cite{Musakhanov:1998wp,Musakhanov:2002vu,Nam:2007gf,Nam:2010pt,Dorokhov:2000gu,Dorokhov:2002iu}. In ~\cite{Nam:2011hg} we use this
model to calculate the fragmentation functions and the quark-distribution functions of the pions. There we limited ourselves in the SU(2) flavor sector. Our result is substantially different from the results of other models with local quark-meson couplings.
In the present work, we extend our studies into the three flavors, $(u,d,s)$, i.e. those functions for the positively charged pion and kaon, i.e. $\pi^+$ and $K^+$, taking into account the explicit flavor-SU(3)-symmetry breaking. We present the result with the two different explicit SU(3) flavor breaking patterns associated with two sets of the current quark masses. 
Furthermore, we figure out that our result of kaons fragmentation functions are significantly different from the ones from the other models.The calculated distribution and fragmentation functions are evolved to high $Q^2$ by the DGLAP evolution using the code {\it QCDNUM}~\cite{Botje:2010ay,DGLAP} to compare with the empirical data. By doing so, we see qualitative agreement between our result and the empirical data, although some underestimates are shown in the small-$(x,z)$ regions, according to the absence of the quark-jet contributions possibly~\cite{Ito:2009zc,Matevosyan:2010hh,Matevosyan:2011ey,Matevosyan:2011vj}.

The present report is organized as follows: In Section II, we briefly introduce NLChQM that we use and sketch our computation procedure. In Section III we present our numerical results and related discussions. The final Section is devoted for the conclusion and future perspectives.
\section{Nonlocal pion-quark coupling}
The process that an off-shell quark ($q$) is fragmented into {an unobserved set of particles} ($X$) and on-shell hadron ($h$), i.e. $q\to h\,X$.
is described by the  unpolarized fragmentation function $D^h_{q}$.
\begin{figure}
\includegraphics[width=12cm]{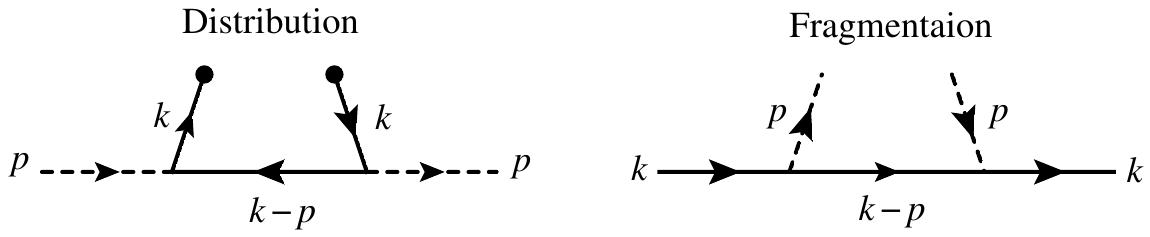}
\caption{Schematic figures for the quark-distribution (left) and fragmentation (right) functions,
in which the solid and dash lines denote the quark and pseudoscalar meson, respectively.}
\label{FIG0}
\end{figure}
A schematic figure for this function is given in the right of Fig.~\ref{FIG0}.
It can be written with the light-cone gauge as follows~\cite{Bacchetta:2002tk,Amrath:2005gv}:
\begin{equation}
\label{eq:FRAG}
D^h_{q}(z,\bm{k}^2_T,\mu)=\frac{\mathcal{C}^h_{q}}{4z}\int dk^+\mathrm{Tr}
\left[\Delta(k,p,\mu)\gamma^- \right]|_{zk^-=p^-},
\end{equation}
where Note that $\mathcal{C}^h_{q}$ indicates the flavor factor for the corresponding fragmentation processes.
The values for $\mathcal{C}^h_{q}$ for the PS meson, i.e. $h=\phi=(\pi,K)$, are given in Table~\ref{TABLE0} in Appendix. $k$, $p$,
and $z$ indicate the four-momenta of the initial quark and fragmented hadron, and the longitudinal momentum fraction possessed by the hadron, respectively.
$k_{\pm}$ denotes $(k_{0}\pm k_{3})/\sqrt{2}$ in the light-cone coordinate.
All the calculations are carried out in the frame where $\bm{k}_{\perp}=0$. Here the $z$-axis is chosen to be the direction of $\bm{k}$. On the other hand,
$\bm{k}_{T}=\bm{k}-[(\bm{k}\cdot\bm{p})/|\bm{p}|^2]\,\bm{p}$, defined as the transverse momentum of the initial quark with respect to the
direction of the momentum of the produced hadron, is nonzero. $\mu$ denotes the renormalization scale at which the fragmentation process computed.
Note that we consider this renormalization scale is almost the same with the momentum-transfer scale, i.e. $\mu^2\approx Q^2$ for simplicity, unless otherwise stated.
The correlation $\Delta(k,p,\mu)$ reads generically:
\begin{equation}
\label{eq:COR1}
\Delta(k,p,\mu)=\sum_X\int\frac{d^4\xi}{(2\pi)^4}e^{+ik\cdot\xi}
\langle0|\psi(\xi)|h,X\rangle\langle h,X|\psi(0)|0\rangle.
\end{equation}
Here $\psi$ represents the quark field, whereas $\xi$ the spatial interval on the light cone.
Furthermore one can integrate over $\bm{k}_{T}$,
\begin{equation}
D^h_{q}(z,\mu)=\pi z^2\int^{\infty}_{0}d\bm{k}^{2}_T\,D^{h}_{q}(z,\bm{k}_{T},\mu).
\end{equation}
The factor of $z^2$ is due to the fact that the integration is over $\bm{p}_\perp=\bm{p}-[(\bm{k}\cdot \bm{p})/|\bm{k}|^2]\,\bm{k}$,
the transverse momentum of the produced hadron
with respect to the quark direction, and there is a relation between $\bm{p}_{\perp}$ and $\bm{k}_{T}$: $\bm{p}_\perp=-z\bm{k}_{T}$.
The integrated fragmentation function satisfies the momentum sum rule:
\begin{equation}
\label{eq:SUM}
\int^1_0\sum_{h}zD^h_{q}(z,\mu)\,dz=1,
\end{equation}
where $h$ indicates for all the possible hadrons fragmented. Eq.~(\ref{eq:SUM}) means that all of the momentum of the initial quark $q$ is
transferred into the momenta of the fragmented hadrons. From the Drell-Levi-Yan (DLY) relation~\cite{Drell:1969jm,Amrath:2005gv,Bentz:1999gx,Ito:2009zc,
Matevosyan:2011ey}, $D^h_{q}$ can be related to the parton distribution function $f^h_{q}$, provided that there is a proper analytic continuation.
The relation is as follows,
\begin{equation}
\label{eq:DLY}
D^h_{q}(z)=\frac{z}{6}f^h_{q}\left(x\right),\,\,\,\,\mathrm{where}\,\,\,\,x=\frac{1}{z},
\end{equation}
where $x$ denotes the momentum fraction possessed by a parton inside the hadron.
A schematic figure for the quark-distribution function is depicted in the left of Fig.~\ref{FIG0}.

In this article, we use NLChQM to investigate these nonperturbative objects, i.e. fragmentation and parton distribution functions.
This model is motivated from the dilute instanton liquid model~\cite{Diakonov:1985eg,Shuryak:1981ff,Diakonov:1983hh,Diakonov:2002fq,Schafer:1996wv}.
We note that, to date, various nonperturbative QCD properties have been well studied in terms of the instanton vacuum configuration and the results
are comparable with experiments and LQCD simulations~\cite{Nam:2007gf,Nam:2010pt,Musakhanov:1998wp,Musakhanov:2002vu}. In that model, nonperturbative QCD effects
are deciphered by the nontrivial quark-instanton interactions in the dilute instanton ensemble. However this model by nature is defined in
Euclidean space because the (anti)instantons are well defined there by signaling the tunneling between the infinitely degenerate QCD vacua.
Although there have been no rigorous derivation on the analytic continuation from the instanton physics to those in Minkowski one, there are still several
challenging studies which try to apply the idea of the instanton physics to the physical quantities defined properly only in Minkowski space, such as
the light-cone wave function~\cite{Dorokhov:1991nj,Nam:2006sx,Nam:2006au,Praszalowicz:2001pi}. Following those studies we adopt the effective chiral
action (EChA) from NLChQM in Minkowski space as follows:
\begin{equation}
\label{eq:ECA}
\mathcal{S}_\mathrm{eff}[m_q,\phi]=-\mathrm{Sp}\ln\left[i\rlap{/}{\partial}
-\hat{m}_f-\sqrt{M(\loarrow{\partial}^2)}U^{\gamma_5}\sqrt{M(\roarrow{\partial}^2)}\right],
\end{equation}
where $\mathrm{Sp}$ and $\hat{m}_q$ denote the functional trace $\mathrm{Tr}\int d^4x \langle x|\cdots|x\rangle$ over all the
relevant spin spaces and SU(3) current-quark mass matrix, $\mathrm{diag}(m_u,m_d,m_s)$, respectively. Note that, in deriving EChA in Eq.~(\ref{eq:ECA}),
we simply replace the Euclidean metric for the (anti)instanton effective chiral action into that for Minkowski space. The momentum-dependent effective quark
mass generated from the interactions between the quarks and nonperturbative QCD vacuum, can be written in a simple $n$-pole type form factor as
follows~\cite{Dorokhov:1991nj,Nam:2006sx,Nam:2006au,Praszalowicz:2001pi}:
\begin{equation}
\label{eq:MDM}
M(\partial^2)=M_0\left[\frac{n\Lambda^2}{n\Lambda^2-\partial^2+i\epsilon} \right]^{n},
\end{equation}
where $n$ indicates a positive integer number. We will choose $n=2$ as in the instanton model~\cite{Nam:2006sx,Nam:2006au}.
$\Lambda$ stands for the model renormalization scale. It is related to the average (anti)instanton size $\bar{\rho}$ in principle.
The nonlinear PS-meson field, i.e. $U^{\gamma_5}$ takes a simple form~\cite{Diakonov:2002fq} with the normalization following Ref.~\cite{Bacchetta:2002tk}
to be consistent with the definition of the fragmentation function in Eq.~(\ref{eq:FRAG}):
\begin{equation}
\label{eq:CHIRALFIELD}
U^{\gamma_5}(\phi)=
\exp\left[\frac{i\gamma_{5}(\bm{\lambda}\cdot\bm{\phi})}{2F_{\phi}}\right]
=1+\frac{i\gamma_{5}(\bm{\lambda}\cdot\bm{\phi})}{2F_{\phi}}
-\frac{(\bm{\lambda}\cdot\bm{\phi})^{2}}{8F^{2}_{\phi}}+\cdots,
\end{equation}
where $F_\phi$ and $\lambda^a$ stand for the weak-decay constant for the PS meson $\phi$ and the Gell-Mann matrix.
We note that, however, the value of $F_\pi$ can be determined rather phenomenologically to reproduce relevant physical
quantities and conditions, even in NLChQM, and will discuss this in detail in Section III. We also write explicitly the flavor SU(3) octet PS-meson fields:
\begin{equation}
\label{eq:PHI}
\bm{\lambda}\cdot\bm{\phi}=\sqrt{2}\left(
\begin{array}{ccc}
\frac{1}{\sqrt{2}}\pi^{0}+\frac{1}{\sqrt{6}}\eta&\pi^{+}&K^+\\
\pi^-&-\pi^{0}+\frac{1}{\sqrt{6}}\eta&K^0\\
K^-&\bar{K}^0&-\frac{2}{\sqrt{6}}\eta\\
\end{array} \right).
\end{equation}
By expanding the nonlinear PS-meson field up to $\mathcal{O}(\phi^1)$ from EChA in Eq.~(\ref{eq:ECA}),
one can derive the following effective interaction Lagrangian density {in the coordinate space} for the nonlocal quark-quark-PS meson vertex:
\begin{equation}
\label{eq:LAG}
\mathcal{L}_{qq\phi}=\frac{i}{2F_\phi}\bar{q}\sqrt{M(\loarrow{\partial}^2)}
\gamma_5(\bm{\tau}\cdot\bm{\phi})\sqrt{M(\roarrow{\partial}^2)}q.
\end{equation}
As we have done in our work in SU(2) sector ~\cite{Nam:2011hg},
we reach a concise expression for the elementary fragmentation function $q\to\phi q'$ from NLChQM:
\begin{equation}
\label{eq:FRAGDEF}
d^{\phi}_{q}(z,\bm{k}^2_T,\mu)=\frac{\mathcal{C}^\phi_{q}}{8\pi^3z(1-z)}\frac{M_kM_{r}}{2F^2_\phi}
\left[\frac{z(k^2-\bar{M}^2_f)+(k^2+\bar{M}^2_f-2\bar{M}_f\bar{M}_{f'}-2k\cdot p) }{(k^2-\bar{M}^2_f)^2}\right].
\end{equation}
here the momentum dependent effective quark mass is given by
\begin{equation}
\label{eq:MDQM}
M_\ell=M_0\left[\frac{2\Lambda^2}{2\Lambda^2-\ell^2+i\epsilon} \right]^{2}.
\end{equation}
and some notations are defined as follows:
$\bar{M}_\ell\equiv m_f+M_\ell$ and $\bar{M}_f\equiv m_f+M_0$. Note that $f$ and $f'$
indicate the flavors for the initial $(q)$ and final $(q')$ quarks, respectively. Note that the value of $M_0$ can be fixed self-consistently within the instanton model~\cite{Diakonov:1985eg,Shuryak:1981ff,Diakonov:1983hh,
Schafer:1996wv,Musakhanov:1998wp,Musakhanov:2002vu,Diakonov:2002fq,Nam:2007gf,Nam:2010pt} with the phenomenological (anti)instanton parameters $\bar{\rho}\approx1/3$ fm and
$\bar{R}\approx1$ fm, resulting in $M_0\approx350$ MeV by the following self-consistent equation, defined in Euclidean ($E$) space. The masses for the pion and kaon are chosen to be $m_{\pi,K}=(140,495)$ MeV throughout the present work.
Collecting all the ingredients above, one is led to a final expression for the elementary fragmentation function:
\begin{equation}
\label{eq:DDDDD}
d^{\phi}_{q}(z,\bm{k}^2_T,\mu)=\frac{\mathcal{C}^\phi_{q}}
{8\pi^3}\frac{M_kM_{r}}{2F^2_\phi}
\frac{z\left[z^2\bm{k}^2_T+[(z-1)\bar{M}_f+\bar{M}_{f'}]^2\right]}
{[z^2\bm{k}^2_T+z(z-1)\bar{M}^2_f+z\bar{M}^2_{f'}+(1-z)m^2_\phi]^2},
\end{equation}
where $M_k$ and $M_{r}$ are the momentum-dependent quark mass manifesting the nonlocal quark-PS meson interactions, read:
\begin{equation}
\label{eq:MASS}
M_k=\frac{M_0[2\Lambda^2z(1-z)]^2}
{[z^2\bm{k}^2_T+z(z-1)(2\Lambda^2-\delta^2)+z\bar{M}^2_{f'}+(1-z)m^2_\phi]^2},
\,\,\,\,
M_{r}=\frac{M_0(2\Lambda^2)^2}{(2\Lambda^2-\bar{M}^2_{f'})^2}.
\end{equation}
As in ~\cite{Nam:2011hg}, a free and finite-valued parameter $\delta$ in the denominator to avoid the unphysical singularities
which appear in the vicinity of $(z,\bm{k}_T)=0$, due the present parametrization of the effective quark mass as in Eq.~(\ref{eq:MDM}). At the renormalization scale in our model, the elementary fragmentation function is assumed to be able to be evaluated further by integrating Eq.~(\ref{eq:DDDDD}) over $k_T$ as:
\begin{equation}
\label{eq:FRAGINT}
d^{\phi}_{q}(z,\mu)=2\pi z^2\int^\infty_0 d^{\phi}_{q}(z,\bm{k}^2_T,\mu)\,\bm{k}_T\,d\bm{k}_T,
\end{equation}
where the factor $z^2$ in the right-hand-side is again comes from the integration over $\bm{p}_\perp=-\bm{k}_T/z$.
Actually the connection between $d^{\phi}_{q}(z,\mu)$ and $d^{\phi}_{q}(z,\bm{k}^2_T,\mu)$ would be far more complicated
in principle~\cite{Aybat:2011zv}. We will present the numerical results for Eq.~(\ref{eq:FRAGINT}) in the next Section.

Using the DLY relation in Eq.~(\ref{eq:DLY}), one can derive the quark-distribution function $f^{\phi}_{q}$ as follows:
\begin{equation}
\label{eq:PDF0}
f^{\phi}_{q}(x,\bm{k}^2_T,\mu)=\frac{3\mathcal{C}^\phi_{q}}{4\pi^3}
\frac{\mathcal{M}_k\mathcal{M}_{r}}{2F^2_\phi}
\frac{\bm{k}^2_T+[(x-1)\bar{M}_f-x\bar{M}_{f'}]^2}
{[\bm{k}^2_T+(1-x)\bar{M}^2_f+x\bar{M}^2_{f'}+x(x-1)m^2_\phi]^2},
\end{equation}
where we have defined the effective quark masses for the quark distribution function by
\begin{equation}
\label{eq:EFFFFF}
\mathcal{M}_k=\frac{4M_0\Lambda^4(1-x)^2}
{[\bm{k}^2_T+(1-x)(2\Lambda^2)+x\bar{M}^2_{f'}+x(x-1)m^2_\phi]^2},\,\,\,\,
\mathcal{M}_r=\frac{4M_0\Lambda^4}{(2\Lambda^2-\bar{M}^2_{f'})^2}.
\end{equation}
We see that Eq.~(\ref{eq:PDF0}) is equivalent to that given Ref.~\cite{Matevosyan:2010hh}, except for the momentum-dependent effective quark mass and $qq\phi$ coupling. Note that $\mathcal{M}_k$ in Eq.~(\ref{eq:EFFFFF}) does not suffer from the unphysical singularities unlike that in Eq.~(\ref{eq:MASS}), according to the different kinematic situations between those functions. Also it ensures the correct behavior that the quark-distribution function becomes zero as $x\to1$~\cite{Shigetani:1993dx}. 

Similarly, the integration over $k_T$ can be performed as follows, resulting in a function of $x$ at a certain renormalization scale $\mu\approx\Lambda$:
\begin{equation}
\label{eq:PDF2}
f^{\phi}_{q}(x,\mu)
=2\pi\int^\infty_0f^{\phi}_{q}(x,\bm{k}^2_T,\mu)\,\bm{k}_T\,d\bm{k}_T.
\end{equation}
The minus-type quark-distribution function, which is nothing but the valance-quark distribution, satisfies the following normalization condition:
\begin{equation}
\label{eq:PDFNORM}
\int^1_0 dx\,\left[f^{\phi}_q(x,\mu)-f^{\phi}_{\bar{q}}(x,\mu) \right]=n_q,
\end{equation}
where $n_q$ in the right-hand-side indicates the valance-quark number and becomes unity. We will tune the $F_\Phi$ value to fulfill the normalization condition in Eq.~(\ref{eq:PDFNORM}) in the next Section.

\section{Result and Discussion}
\begin{figure}[t]
\includegraphics[width=8.5cm]{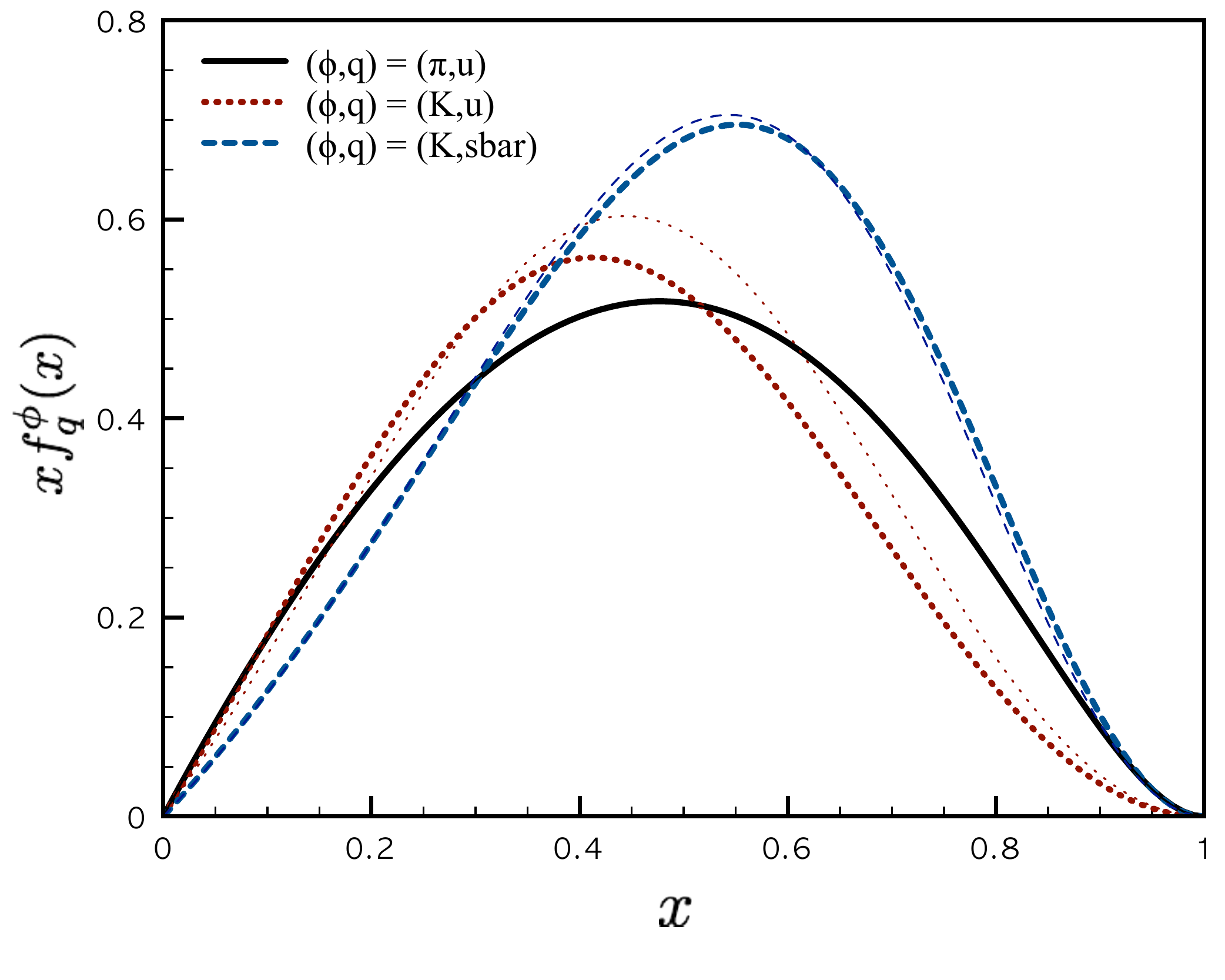}
\caption{(Color online) Quark-distribution functions, multiplied by $x$, i.e. $xf^\phi_q(x)$ for $(\phi,q)=(\pi,u)$ (solid),
$(\phi,q)=(K,u)$ (dot), and $(\phi,q)=(K,\bar{s})$ (dash), at $Q^2=\Lambda^2=0.36\,\mathrm{GeV}^2$. The thick and thin lines indicate the results from Model I and II, respectively. The curves for $xf^{\pi}_u(x)$ for Model I and II show negligible differences as shown.}
\label{FIG1}
\end{figure}
In this section, we present and discuss our numerical results. We only compute the fragmentation functions for $\pi^+$ and $K^+$ because the other PS-meson
channels can be easily obtained by multiplying the flavor factors in Table~\ref{TABLE0}. It is due to the fact that we
are only focusing on the elementary process here.
For simplicity, we will denote the positively charged pion and kaon simply as $\pi$ and $K$ hereafter.
The values of $F_{\pi,K}$ in Eqs.~(\ref{eq:DDDDD}) and (\ref{eq:PDF0})
are determined by the normalization conditions for the quark distribution functions as in Eq.~(\ref{eq:PDFNORM}).
In this phenomenological approach, we take into account two models with different patterns of the explicit flavor SU(3) breaking
according to their current-quark masses:
\begin{itemize}
\item {\it Model I}\\
We consider the isospin symmetry for the $u$ and $d$ quarks, and  $m_s\sim\Lambda_\mathrm{QCD}$: $(m_u,m_d,m_s)=(5,5,150)$ MeV.
\begin{equation}
\label{eq:WDC1}
F^{u\to\pi d}_\pi=88.07\,\mathrm{MeV},\,\,\,\,
F^{u\to K s}_K=118.30\,\mathrm{MeV},\,\,\,\,
F^{\bar{s}\to K\bar{u}}_K=103.20\,\mathrm{MeV}.
\end{equation}
\item {\it Model II}\\
We employ the current-quark masses from the PDG center values~\cite{Nakamura:2010zzi}: $(m_u,m_d,m_s)=(2.5,5,100)$ MeV.
\begin{equation}
\label{eq:WDC2}
F^{u\to\pi d}_\pi=88.49\,\mathrm{MeV},\,\,\,\,
F^{u\to Ks}_K=116.60\,\mathrm{MeV},\,\,\,\,
F^{\bar{s}\to K\bar{u}}_K=95.94\,\mathrm{MeV}.
\end{equation}
\end{itemize}
Note that $F^{u\to\pi d}_\pi$ and $F^{\bar{s}\to K\bar{u}}_K$ are about $10\%$ smaller than their empirical values, $F_{\pi,K}\approx(93,113)$ MeV.
This has already been observed in Refs.~\cite{Nam:2011hg,Nam:2008bq} and explained by the fact that the so-called nonlocal contributions proportional to
$\partial M(p^2)/\partial |p|$ and the meson-loop contributions corresponding to large-$N_c$ corrections have been neglected.
Due to the similar reason, $F^{u\to Ks}_K$ turns out to be slightly larger than its empirical value. Instead of including the nonlocal contributions in the present work, we have modified the $F_{\pi,K}$ values as in Eqs.~(\ref{eq:WDC1}) and (\ref{eq:WDC2}),
satisfying the normalization condition for $f^\phi_q(x)$ for brevity~\cite{Nam:2011hg}.
This treatment is supposed to compensate the absence of the nonlocal contributions phenomenologically.
\begin{figure}
\begin{tabular}{cc}
\includegraphics[width=8.5cm]{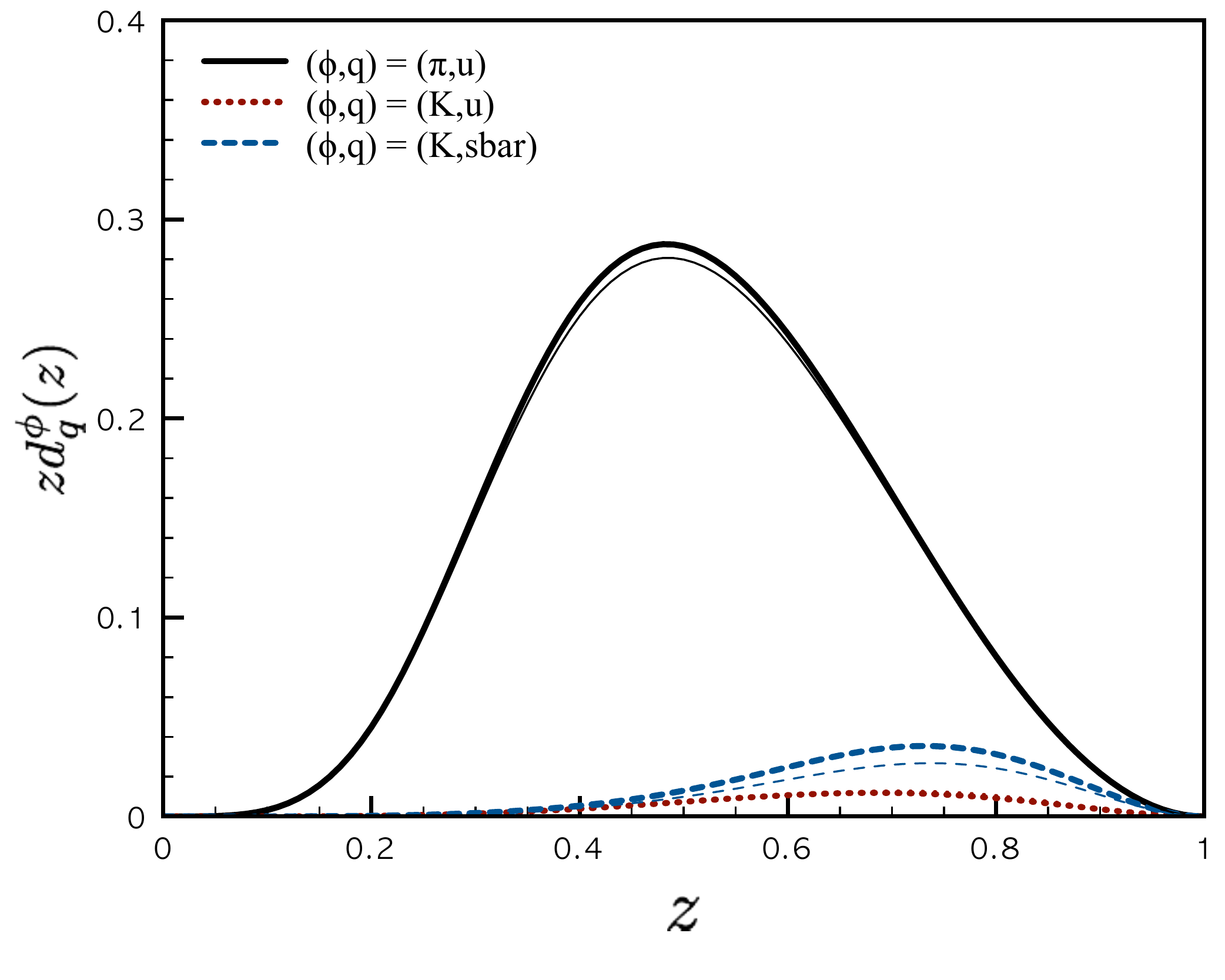}
\includegraphics[width=8.5cm]{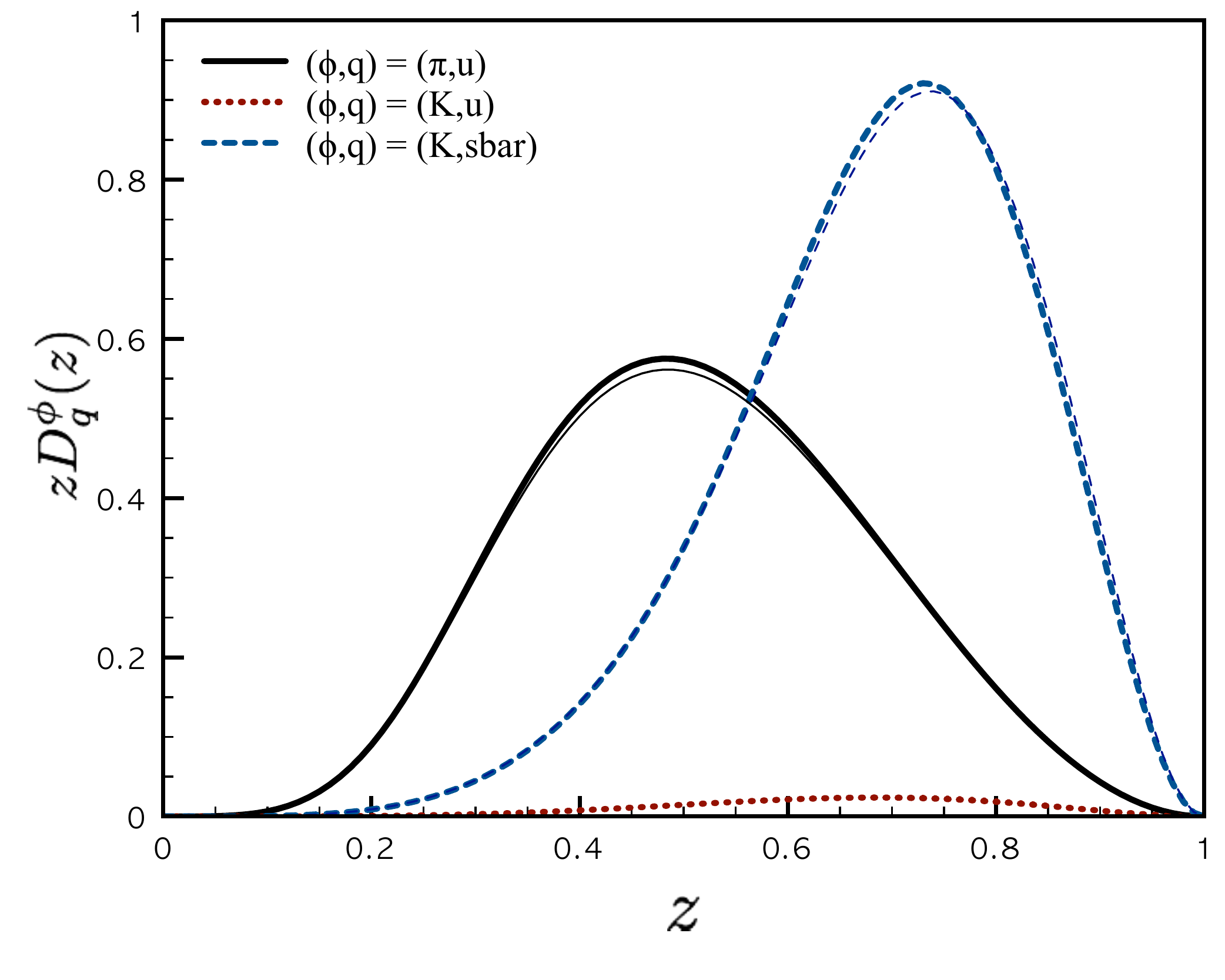}
\end{tabular}
\caption{(Color online) Elementary (left) and renormalized (right) fragmentation functions, multiplied by $z$, i.e. $zd^\phi_q(z)$ and $zD^\phi_q(z)$, at $Q^2=\Lambda^2=0.36\,\mathrm{GeV}^2$. The thick and thin lines indicate the results from Model I and II, respectively.}
\label{FIG23}
\end{figure}
\begin{figure}
\begin{tabular}{cc}
\includegraphics[width=8.5cm]{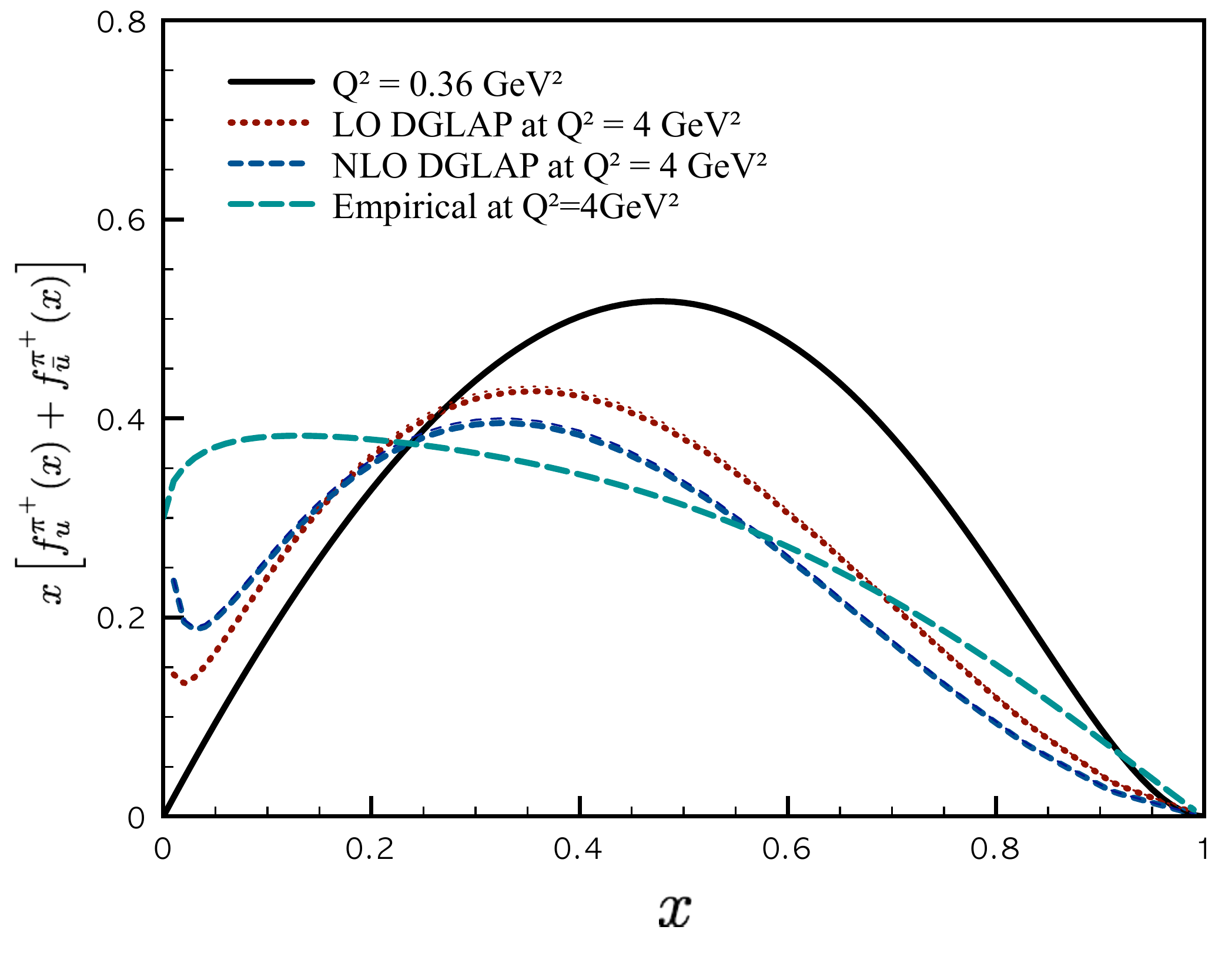}
\includegraphics[width=8.5cm]{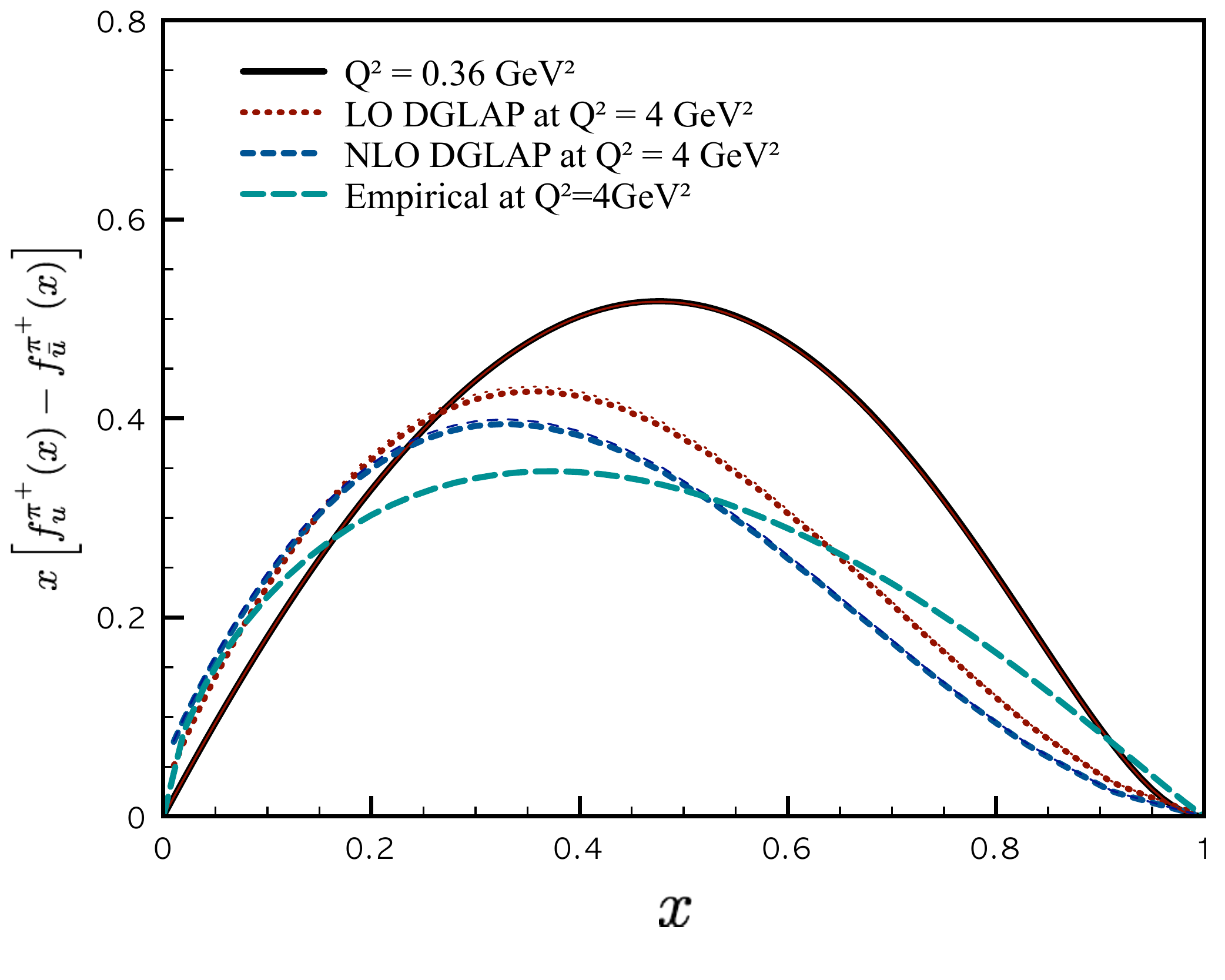}
\end{tabular}
\begin{tabular}{cc}
\includegraphics[width=8.5cm]{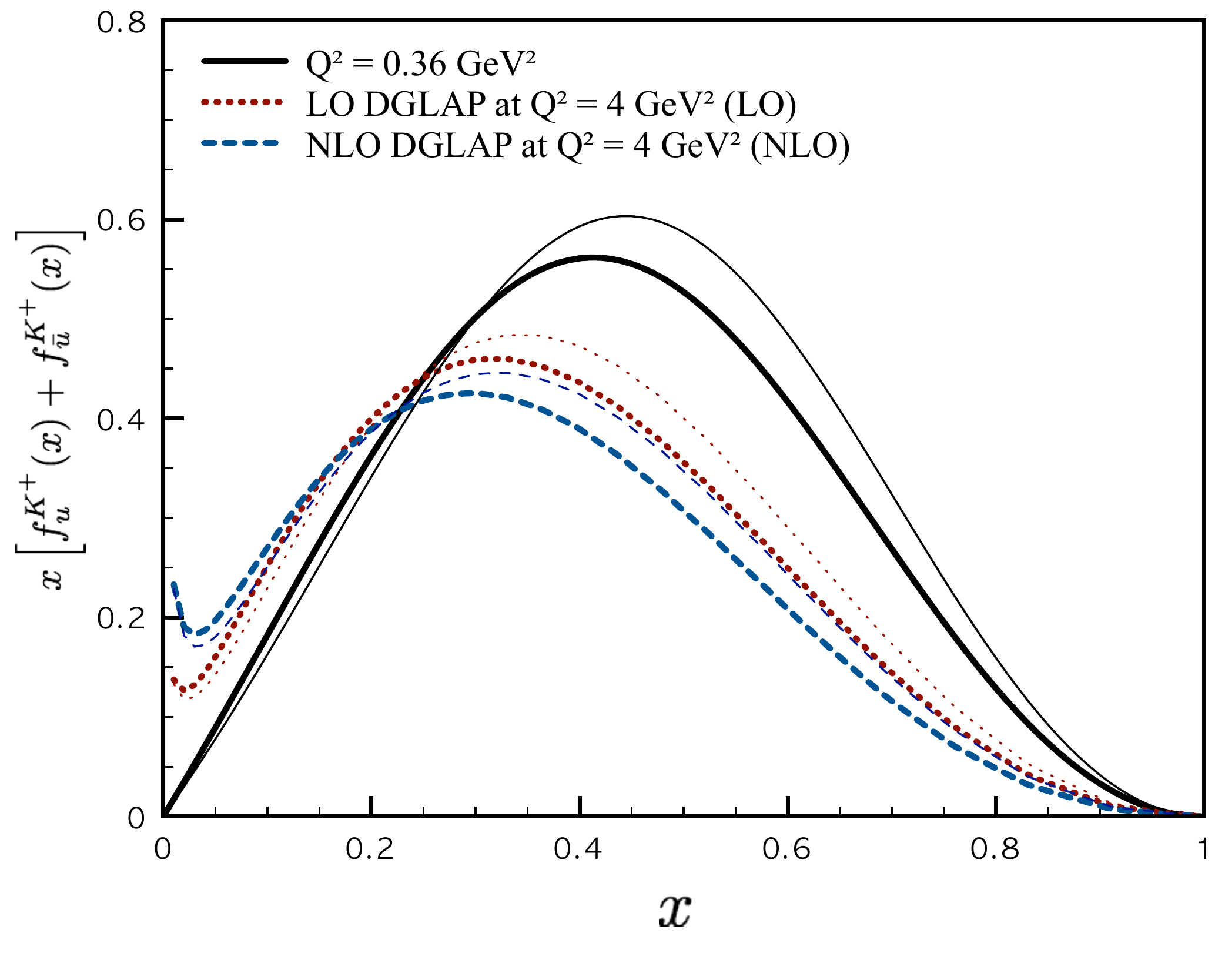}
\includegraphics[width=8.5cm]{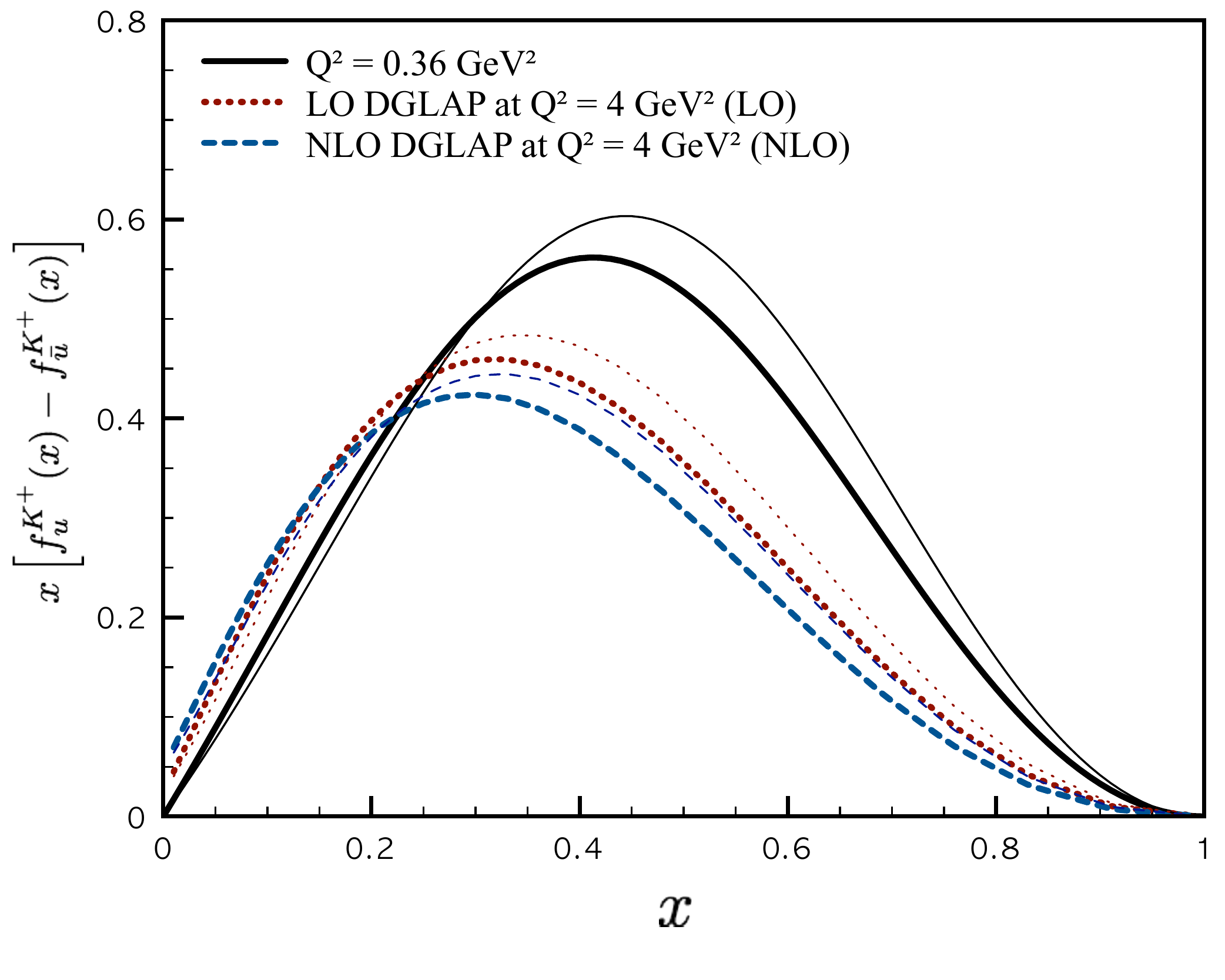}
\end{tabular}
\begin{tabular}{cc}
\includegraphics[width=8.5cm]{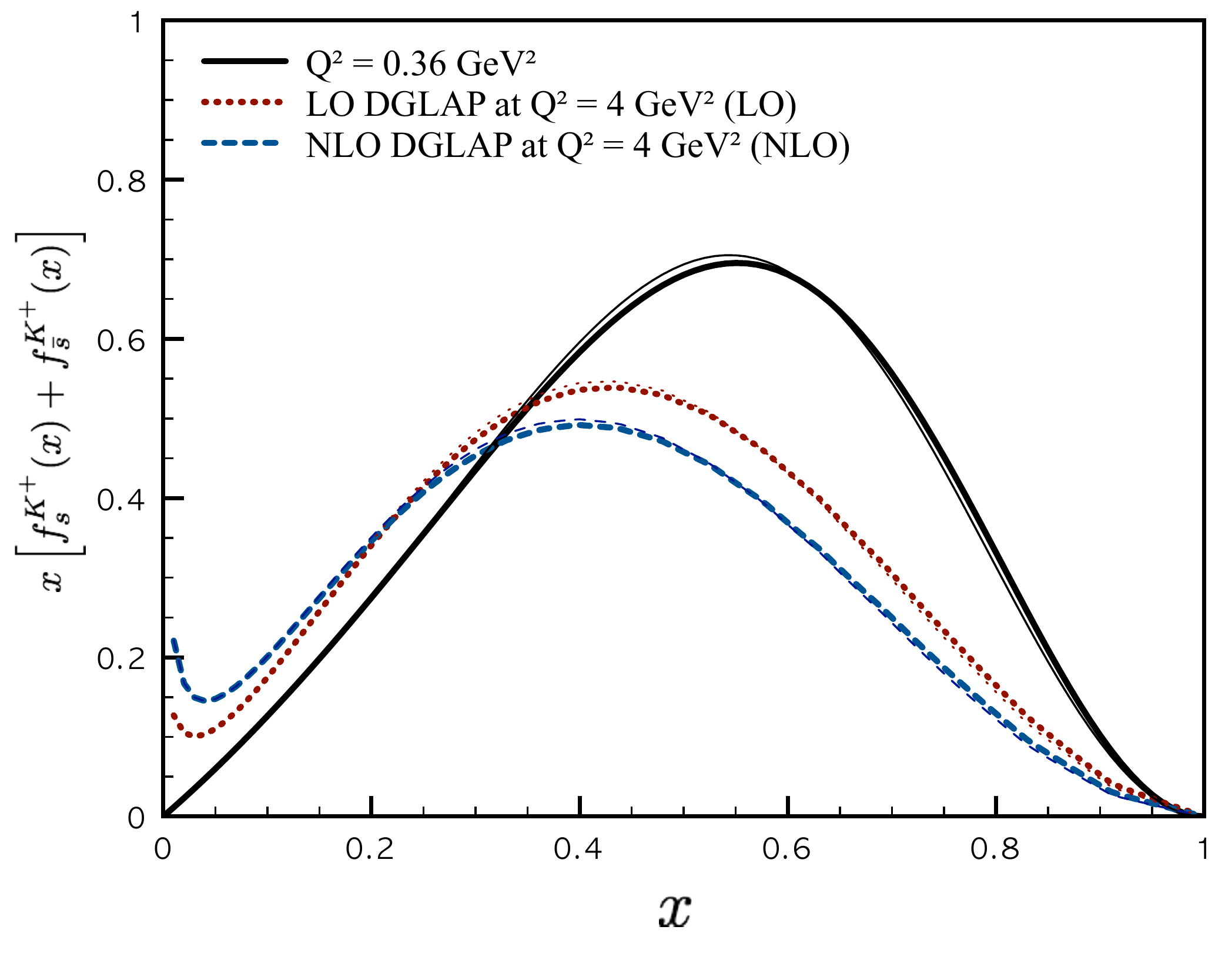}
\includegraphics[width=8.5cm]{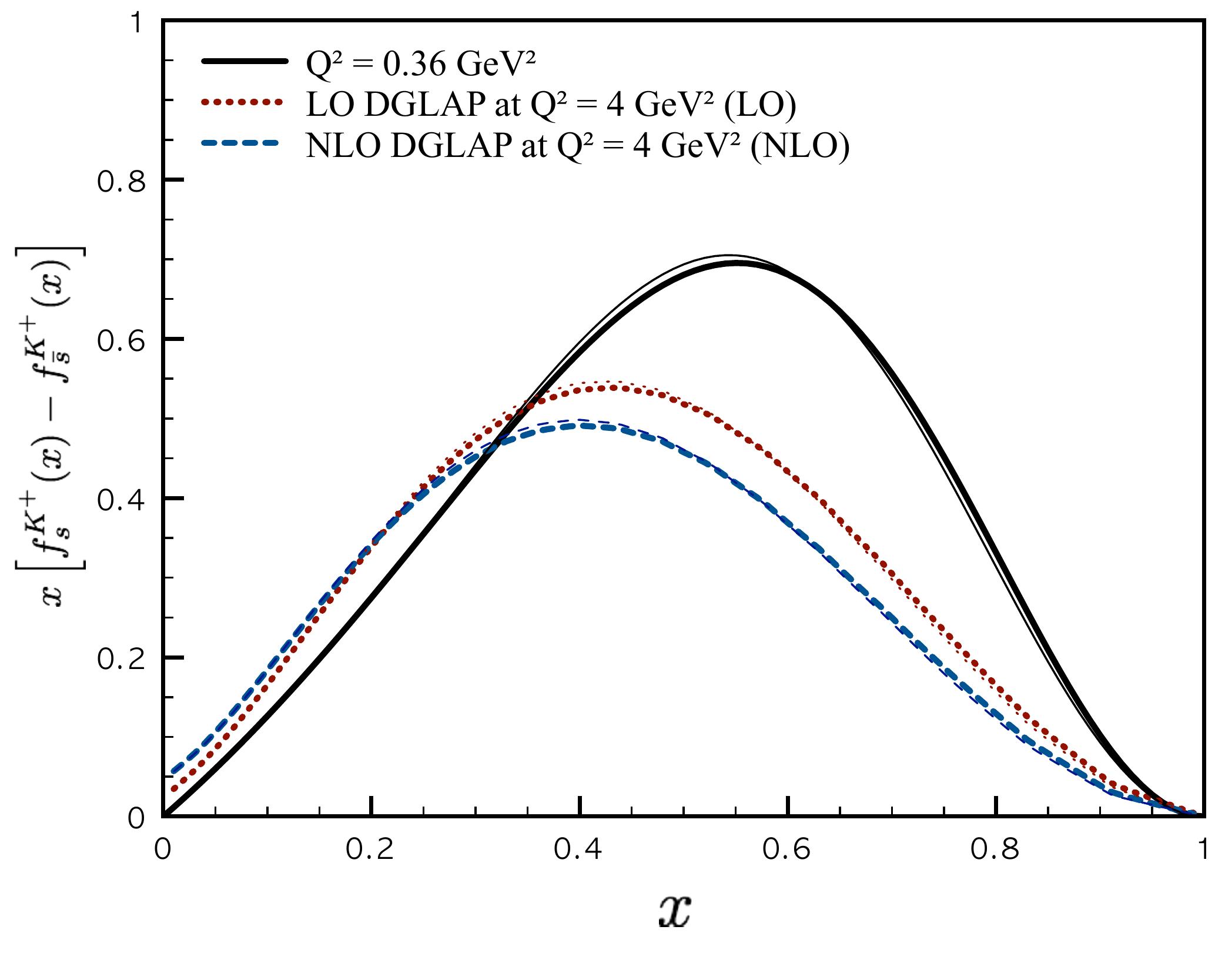}
\end{tabular}
\caption{(Color online) Plus-type (left) and min
us-type (right) quark-distribution functions, multiplied by $x$, i.e. $x[f^\phi_q(x)\pm f^{\phi}_{\bar{q}}(x)]$ for $\phi=(\pi^+,K^+)$ and $q=(u,s)$, at $Q^2=\Lambda^2=0.36\,\mathrm{GeV}^2$ (solid). The dot and dash lines indicate the LO and NLO DGLAP evolved
functions, respectively, at $Q^2=4\,\mathrm{GeV}^2$. The thick and thin lines indicate the results from Model I and II, respectively.}
\label{FIG45}
\end{figure}
Using the values in Eqs.~(\ref{eq:WDC1}) and (\ref{eq:WDC2}), we first present the numerical results for the quark-distribution functions,
multiplied by $x$ in Fig.~\ref{FIG1}, i.e. $xf^\phi_q(x)$ for $(\phi,q)=(\pi,u)$ (solid), $(\phi,q)=(K,u)$ (dotted), and $(\phi,q)=(K,\bar{s})$ (dashed),
at $Q^2=\Lambda^2=0.36\,\mathrm{GeV}^2$. The thick and thin lines represent the results from Model I and II, respectively. We find that the curves of
$xf^\pi_u(x)$ from Model I and II are not distinguishable in the given resolution. It means the isospin-symmetry breaking effects are negligible as expected.
Similarly the curves of $xf^K_{\bar{s}}(x)$ from different models show very small difference. In contrast, there is a clear
difference between the two curves of $xf^K_{u}(x)$ in Fig.~\ref{FIG1}.
In the other words, the fraction of momentum of the $u$ quark is more sensitive to the $s$ quark-mass than 
the one of the $\bar{s}$ quark inside the kaon. 
In terms of shape, we obtain very symmetric curves for $xf^\pi_u(x)$ around $x=1/2$, whereas $xf^K_{u,\bar{s}}(x)$ are both tilted slightly 
to one side.
For other applications such as the DGLAP evolution via QCDNUM17, which is a FORTRAN program that numerically evolves parton densities or
fragmentation functions up to NNLO in perturbative QCD~\cite{Botje:2010ay,DGLAP}, it is very useful to parameterize the quark-distribution functions in the form as
\begin{equation}
\label{eq:PARA}
xf^\phi_q(x)=A^\phi_q\,x^{\alpha^\phi_q}\,(1-x)^{\beta^\phi_q},
\end{equation}
where $A^\phi_q$, $\alpha^\phi_q$, and $\beta^\phi_q$ denote real constants. Using Eq.~(\ref{eq:PARA}),
we can parameterize the present numerical results shown in Fig.~\ref{FIG1} as follows:
\begin{itemize}
\item Model I
\begin{equation}
\label{eq:PARA1}
xf^{\pi}_u(x)=3.22\,x^{1.18}\,(1-x)^{1.43},\,\,\,\,
xf^{K}_u(x)=7.08\,x^{1.48}\,(1-x)^{2.25},\,\,\,\,
xf^{K}_{\bar{s}}(x)=8.77\,x^{1.86}\,(1-x)^{1.74}.
\end{equation}
\item Model II
\begin{equation}
\label{eq:PARA2}
xf^{\pi}_u(x)=3.21\,x^{1.18}\,(1-x)^{1.42},\,\,\,\,
xf^{K}_u(x)=8.51\,x^{1.62}\,(1-x)^{2.21},\,\,\,\,
xf^{K}_{\bar{s}}(x)=9.97\,x^{1.92}\,(1-x)^{1.84}.
\end{equation}
\end{itemize}
After fixing $F_{\pi,K}$ for Model I and II, now we are in a position to present the fragmentation functions in Eq.~(\ref{eq:DDDDD}). In the left panel of Fig.~\ref{FIG23}, we show the numerical results for the {\it elementary} fragmentation functions, multiplied by
$z$, i.e. $zd^\phi_q(z)$, at $Q^2=\Lambda^2=0.36\,\mathrm{GeV}^2$, in the same manner with Fig.~\ref{FIG1}. The thick and thin lines indicate the results from
Model I and II, respectively. We observe that the pion fragmentation is much larger than those of the kaon. Furthermore, the curves of $zd^{K}_{u}$ and $zd^{\pi}_{\bar{s}}$ are titled to the region $z\gtrsim0.5$.
The differences between Model I and II for are all not significant in the case of fragmentation functions.

In Refs.~\cite{Bentz:1999gx,Ito:2009zc}, it was argued that the elementary fragmentation functions
do not hold the momentum sum rule in Eq.~(\ref{eq:SUM}). To remedy this problem, the PS-meson cloud effects were devised and enumerated by the following relation:
\begin{equation}
\label{eq:POP}
\sum_{\phi'}\int^1_0dz\,d^{\phi'}_q(z)\equiv \mathcal{N}_{q},
\end{equation}
where $\mathcal{N}_{q}$ stands for the {\it renormalization} constant representing the PS-meson cloud effects.
The summation runs over all the possible PS-meson states $\phi'$, which can be fragmented from $q$. As for the PS-mesons fragmented
from one $u$ quark, we have to sum over $\phi'=(\pi^+,\pi^0,K^+)$ to obtain $\mathcal{N}_q$. The values of $\mathcal{C}^\phi_q$ are listed in Table~\ref{TABLE0} in Appendix. Note that we do not include $\eta$ and $\eta'$ in $\phi'$ because one needs calculate flavor-singlet contribution if one wants to include them in $\phi'$ and it is beyond our current framework. 

Typical value of $\mathcal{N}$ is estimated as $(0.1\sim0.2)$ in other models. For instance, in according to the NJL model~\cite{Ito:2009zc,
Matevosyan:2010hh,Matevosyan:2011ey,Matevosyan:2011vj} or quark-PS meson coupling models~\cite{Bacchetta:2002tk,Amrath:2005gv,Bacchetta:2007wc},
it was given that $\mathcal{N}^\mathrm{NJL}_{u}=0.03$, $\mathcal{N}^\mathrm{PS}_{u}=0.08$, and $\mathcal{N}^\mathrm{PV}_{u}=0.17$, respectively.
Here the superscripts PS and PV denote the pseudoscalar and pseudovector quark-PS meson couplings. Numerically, we have $\mathcal{N}_u=0.42\,(0.41)$ and $\mathcal{N}_{\bar{s}}=0.04\,(0.03)$ for Model I (II) in the present framework.
Note that our value for $\mathcal{N}_{\bar{s}}$ is underestimated because we ignore the contribution from $d^{\eta,\eta'}_{\bar{s}}(z)$ here.
These values are obviously larger than those in other models. It tells us that
substantial PS-meson cloud effects have already been included in our model as discussed in the previous work~\cite{Nam:2011hg}. Actually the momentum-dependent effective quark mass contains part of the PS-meson cloud effects, since in usual Dyson-Schwinger approaches~\cite{Nguyen:2011jy} the effective quark mass corresponds partially to the dressed quark mass generated by the PS-meson could. Although the origin of the effective mass in our model is the nontrivial interactions between the quarks and (anti)instantons. Nevertheless, one still can argue that both of PS meson cloud and the quark-(anti)instantons interaction are both from the QCD  vacuum. Thus we define a {\it renormalized} fragmentation function for the pion and kaon
as follows~\cite{Nam:2011hg}:
\begin{equation}
\label{eq:RENORM}
D^\phi_q(z)=\frac{ d^\phi_q(z)}{\mathcal{N}_{q}}.
\end{equation}

In the right panel of Fig.~\ref{FIG23}, we show the numerical results for $zD^\phi_q(z)$ via Eq.~(\ref{eq:RENORM})
in the same manner with the left panel of Fig.~\ref{FIG23}. The curve of $zD^{K}_{\bar{s}}(z)$ is enhanced a lot compared
to the curves of $zD^{\pi}_{u}(z)$ and $zD^{K}_{u}(z)$ because $\mathcal{N}_{u}$ is much larger than $\mathcal{N}_{\bar{s}}$.
$zD^{K}_{u}$ is far more smaller than $zD^{\pi}_{u}$ reflects the reality that the energtic $u$ quark is most likely to
be fragmented into pions rather than kaons.
Moreover, we find the tendency  $zD^K_u\ll zD^\pi_u$ for all the $z$ regions. This large imparity is a quite different from the
NJL-Jet model~\cite{Matevosyan:2010hh} where $zD^K_u\sim zD^\pi_u/2$ approximately. Although they were computed at a
different $Q^2$ value from ours, $Q^2=0.2\,\mathrm{GeV}^2$ and their results have included quark-jet contributions which is supposed to be
very important at low $z$ regime. The difference of $zD^K_u/zD^{\pi}_u$ between their results and ours at high $z$ regime is still interesting.
Presumably it is due to the nonlocality of the quark-meson interactions in our model.

To compare our results with the empirical data which are usually extracted from high-energy scattering experiments~\cite{Badier:1980jq},
we need perform the high-$Q^2$ evolution.
To this end, we will employ QCDNUM17~\cite{Conway:1989fs,Botje:2010ay,DGLAP}. As the inputs for QCDNUM17, we use the parameterized
quark-distribution functions in Eqs.~(\ref{eq:PARA1}) and (\ref{eq:PARA2}), taking into account the elementary processes only for the positively charged pion and kaon:
\begin{eqnarray}
\label{eq:INPUT}
x(f^\pi_u-f^\pi_{\bar{u}})&=&xf^\pi_u,
\,\,\,\,x(f^\pi_d-f^\pi_{\bar{d}})=-xf^\pi_{\bar{d}},
\,\,\,\,x(f^\pi_s-f^\pi_{\bar{s}})=0,
\,\,\,\,xf^\pi_{\bar{u}}=0,
\,\,\,\,xf^\pi_{\bar{d}}=xf^\pi_{\bar{d}},
\,\,\,\,xf^\pi_{\bar{s}}=0,
\cr
x(f^K_u-f^K_{\bar{u}})&=&xf^K_u,
\,\,\,\,x(f^K_d-f^K_{\bar{d}})=0,
\,\,\,\,x(f^K_s-f^K_{\bar{s}})=-xf^K_{\bar{s}},
\,\,\,\,xf^K_{\bar{u}}=0,
\,\,\,\,xf^K_{\bar{d}}=0,
\,\,\,\,xf^K_{\bar{s}}=xf^\pi_{\bar{s}},
\end{eqnarray}
where we have assumed that $xf^\pi_u=xf^\pi_{\bar{d}}$, due to the isospin asymmetry. In Fig.~\ref{FIG45}, we show the numerical results for the
plus-type (left) and minus-type (right) quark-distribution functions, multiplied by $x$, i.e. $x[f^\phi_q(x)\pm f^{\phi}_{\bar{q}}(x)]$ for
$(\phi,q)=(\pi,u)$ (first row), $(\phi,q)=(K,u)$ (second row), and $(\phi,q)=(K,\bar{s})$ (third row). We note that the minus-type function is nothing but the valance-quark distribution function. The results at $Q^2=0.36\,\mathrm{GeV}^2$ are given in the solid line, whereas the dotted and dashed lines indicate the LO and
NLO DGLAP evolved functions at $Q^2=4\,\mathrm{GeV}^2$, respectively. The thick and thin lines indicate the results from Model I and II, respectively. The empirical data for the pion, given in the dashed lines, are extracted from a consistent NLO analysis of several high-statistics $\pi^\pm N$ experiments including both Drell-Yan and prompt photon production~\cite{Sutton:1991ay}.

As for the pion case, Model I and II provide almost the same curves as expected. As $Q^2$ increases,
the plus-type functions are tilted to the smaller $x$ region due to the increasing gluon contributions. Although the plus-type function underestimates the empirical data in the vicinity of $x=0$, the minus-type one reproduces the empirical data qualitatively well. The present underestimates in the small $x$ region may be improved by including the quark-jet contributions as in the NJL-Jet model~\cite{Matevosyan:2010hh,Matevosyan:2011ey,Matevosyan:2011vj}. Here we do not consider those contributions. The second and third rows show those functions for the kaon. We see again similar tendency with that observed in the pion case. The $m_s$ dependence becomes most visible for $(\phi,q)=(K,u)$.

\begin{figure}
\begin{tabular}{cc}
\includegraphics[width=8.5cm]{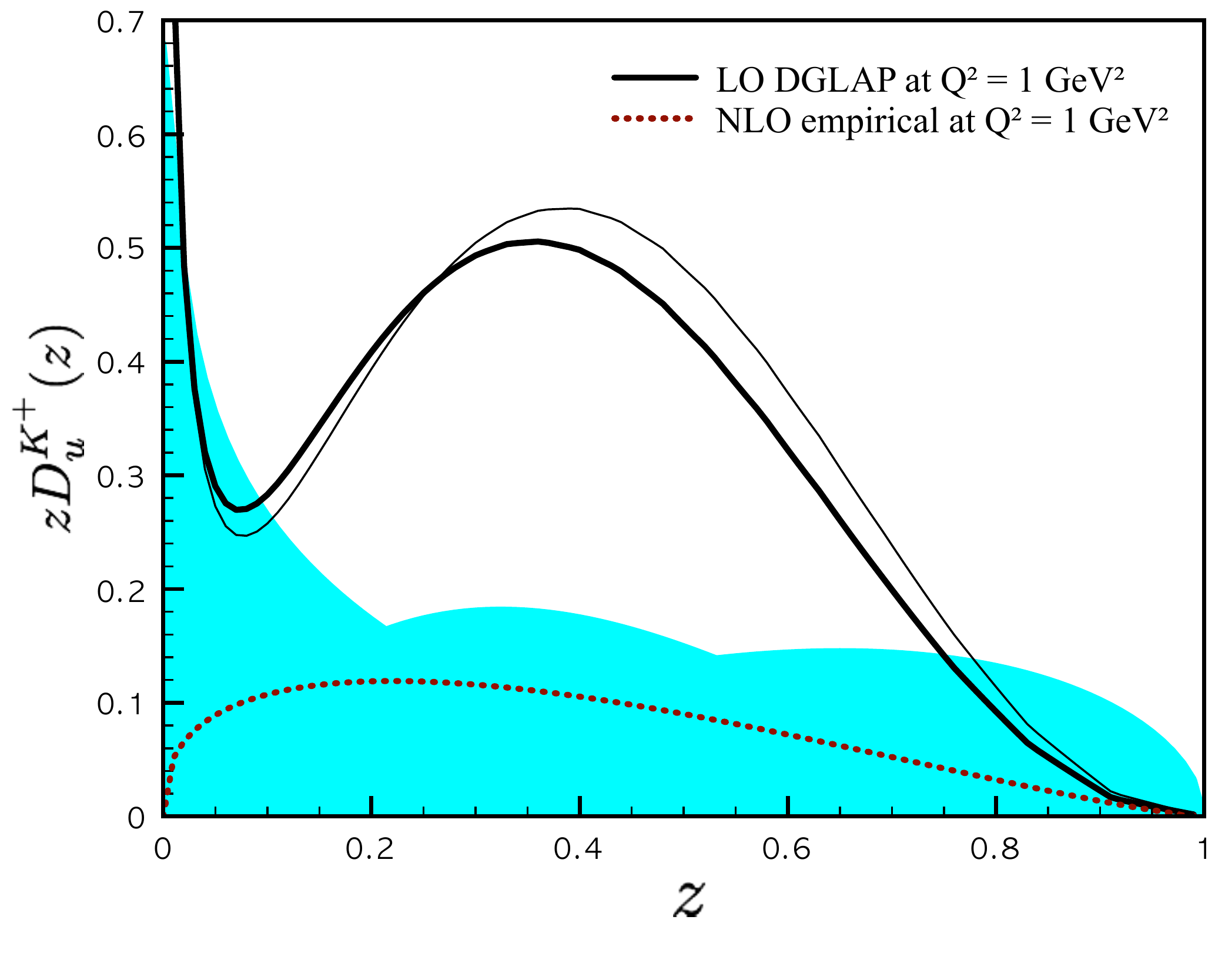}
\includegraphics[width=8.5cm]{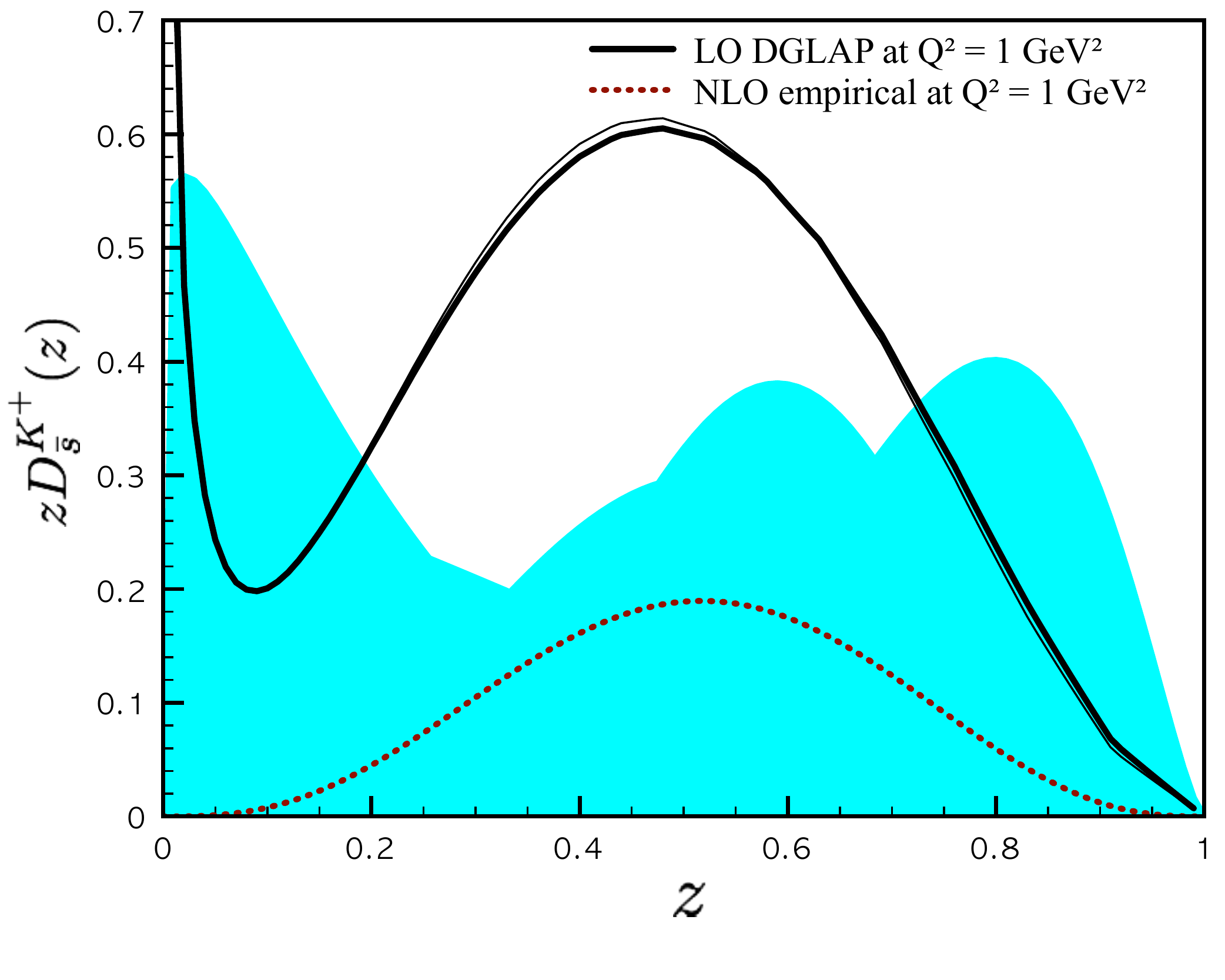}
\end{tabular}
\caption{(Color online) The fragmentation functions $zD^{K^{+}}_{u}$(left) and $zD^{K^{+}}_{\bar{s}}$(right) via LO and NLO DGLAP evolution at $Q^2=1.0\,\mathrm{GeV}^2$. The empirical curves are taken from ~\cite{Hirai:2007cx}.
The thick and thin lines indicate the results from Model I and II, respectively.}
\label{FIG89}
\end{figure}

In Fig.~\ref{FIG89}, we draw the renormalized fragmentation functions, multiplied by $z$ and LO DGLAP evolved to $Q^2=1\,\mathrm{GeV}^2$, for
$(\phi,q)=(K,u)$ (left), and $(\phi,q)=(K,\bar{s})$ (right). The dotted lines represent the empirical curves taken from a global $\chi^2$ analysis of
charged-hadron $(h)$ production data in electron-positron annihilation, i.e. $e^++e^-\to h+X$~\cite{Hirai:2007cx}. The shaded areas indicate the
accumulation of the errors estimated by the Hessian method. In that analysis, the fragmentation functions for the PS mesons are parameterized as follows~\cite{Hirai:2007cx}:
\begin{equation}
\label{eq:FRAGPARA}
zD^\phi_q(z)=\frac{M^\phi_q}{\mathrm{B}(\alpha^\phi_q+1,\beta^\phi_q+2)}\,
z^{\alpha^\phi_q+1}\,(1-z)^{\beta^\phi_q},
\end{equation}
where the NLO parameters, $M^\phi_q$, $\alpha^\phi_q$, and $\beta^\phi_q$ are given in Table~\ref{TABLE1}. $\mathrm{B}(a,b)$ in the denominator stands for
the Beta function with the arguments $a$ and $b$. The values for those NLO parameters are listed in Table~\ref{TABLE1}~\cite{Hirai:2007cx}.
Comparing the empirical data with our results, we observe considerable overshoots around $z=(0.4\sim0.5)$ exceeding the errors.
Moreover, the results from Model I and II are quite close.

\begin{figure}[t]
\begin{tabular}{cc}
\includegraphics[width=8.5cm]{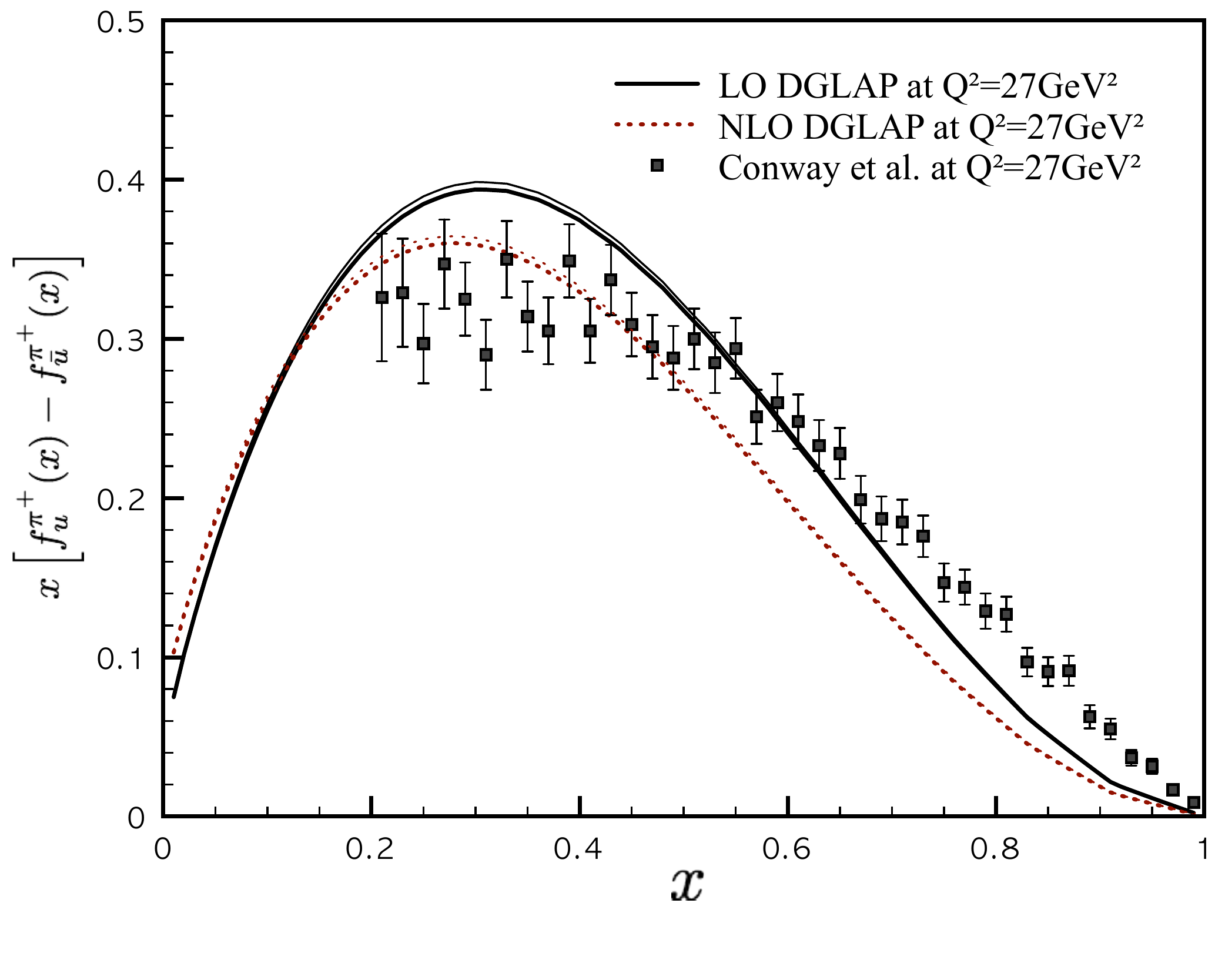}
\includegraphics[width=8.5cm]{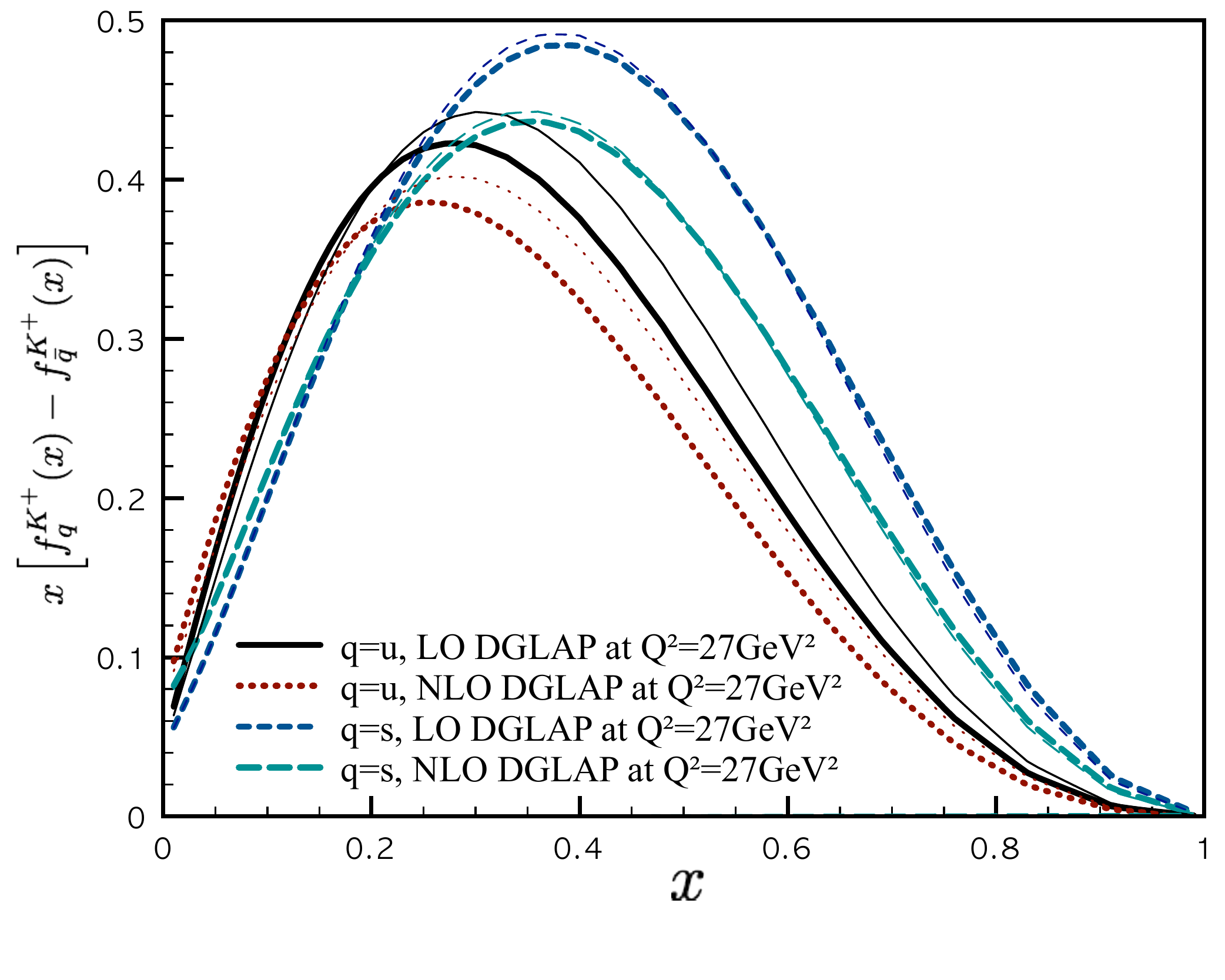}
\label{FIG1213}
\end{tabular}
\caption{(Color online) Valance-quark distribution, $x[f^\phi_q(x)- f^{\phi}_{\bar{q}}(x)]$ for the pion (left) and kaon(right) via LO and NLO DGLAP evolutions at
$Q^2=27\,\mathrm{GeV}^2$. The experimental data of the ion is taken from ~\cite{Conway:1989fs}. The thick and thin lines indicate the results from Model I and II, respectively }
\end{figure}

\begin{table}[b]
\begin{tabular}{c||c|c|c}
&$M^\phi_q$&$\alpha^\phi_q$&$\beta^\phi_q$\\
\hline
\hline
$D^{\pi}_u$&$0.401\pm0.052$&$-0.963\pm0.177$&$1.370\pm0.144$\\
\hline
$D^{K}_u$&$0.0740\pm0.0268$&$-0.630\pm0.629$&$1.310\pm0.772$\\
\hline
$D^{K}_{\bar{s}}$&$0.0878\pm0.0506$&$2.000\pm2.913$&$2.800\pm1.313$\\
\end{tabular}
\caption{Parameters for Eq.~(\ref{eq:FRAGPARA}) at $Q^2=1\,\mathrm{GeV}^2$, taken from a global $\chi^2$ analysis of $e^++e^-\to h+X$~\cite{Hirai:2007cx}.}
\label{TABLE1}
\end{table}
In Fig.~\ref{FIG1213}, we depict the minus-type quark-distribution functions multiplied by $x$, i.e $x[f^\phi_q-f^\phi_{\bar{q}}]$,
evolved to $Q^2=27\,\mathrm{GeV}^2$, for the pion (left) and kaon (right). The LO (solid) and NLO (dotted) data for the pion is taken
from Ref.~\cite{Conway:1989fs} and are extracted from the muon-pair production experiment by $252$ GeV pions on tungsten target.
The Model I and II results are depicted by the thick and thin lines, respectively. As shown in the left panel, the isospin-symmetry breaking is negligible. It turns out that the numerical results reproduce the data qualitatively well.
Note that similar results were already observed in Ref.~\cite{Nam:2011hg}. In the right panel, we show the the kaon case fragmented from $q=u$ for LO (solid) and NLO (dotted) and $q=s$ for LO (dashed) and NLO (long-dashed) in the same manner with the left panel for Model I and II.
The $m_s$ dependencies are obviously shown for the case for $q=u$, being similar to those in Fig.~\ref{FIG45}.
The overall shapes of the curves for the kaon are very similar to those for the pion for the region $x\lesssim0.4$,
but the kaon curves drop more stiffly in comparison to the pion case when $x\ge 0.5$.

\begin{figure}
\begin{tabular}{cc}
\includegraphics[width=8.5cm]{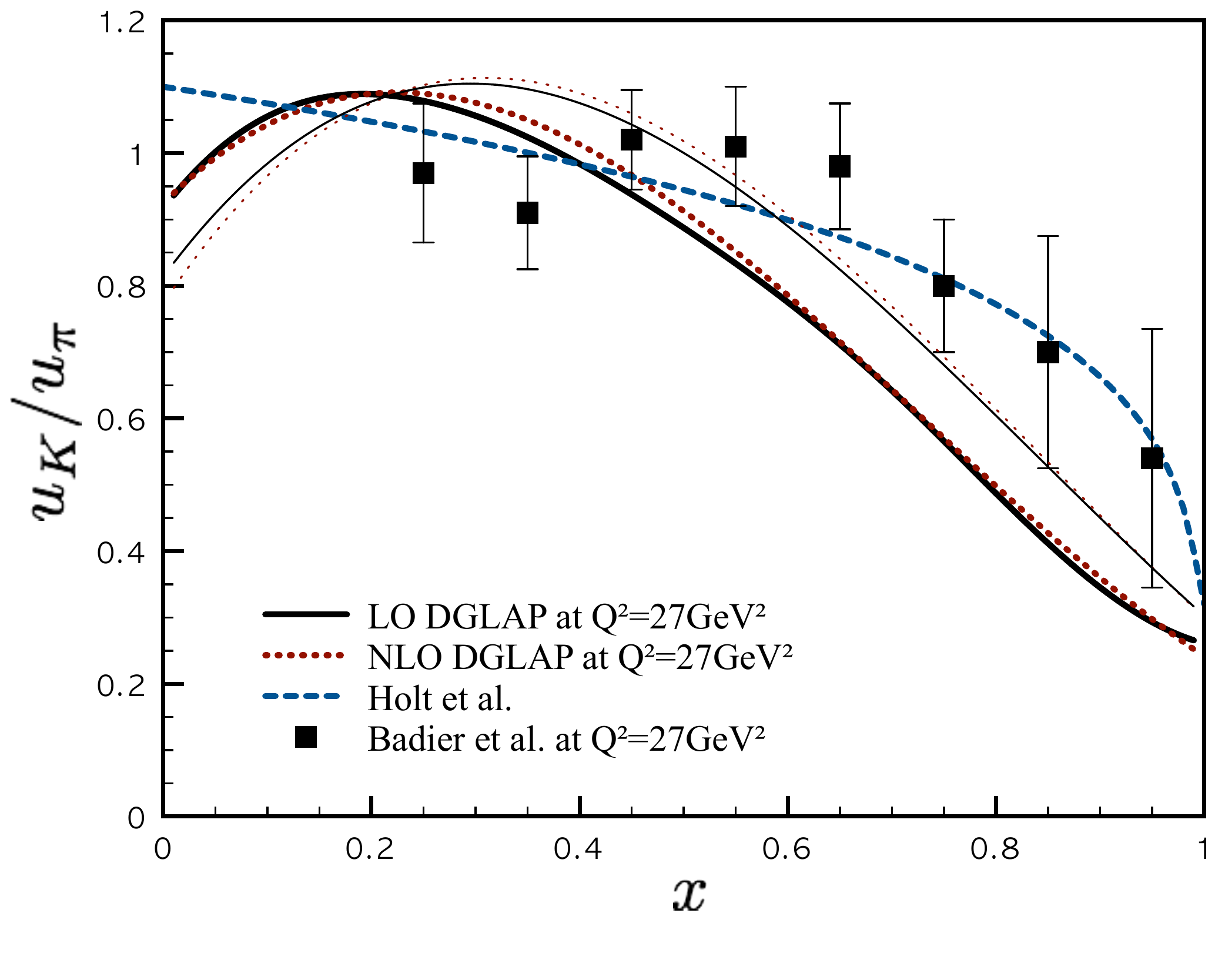}
\includegraphics[width=8.5cm]{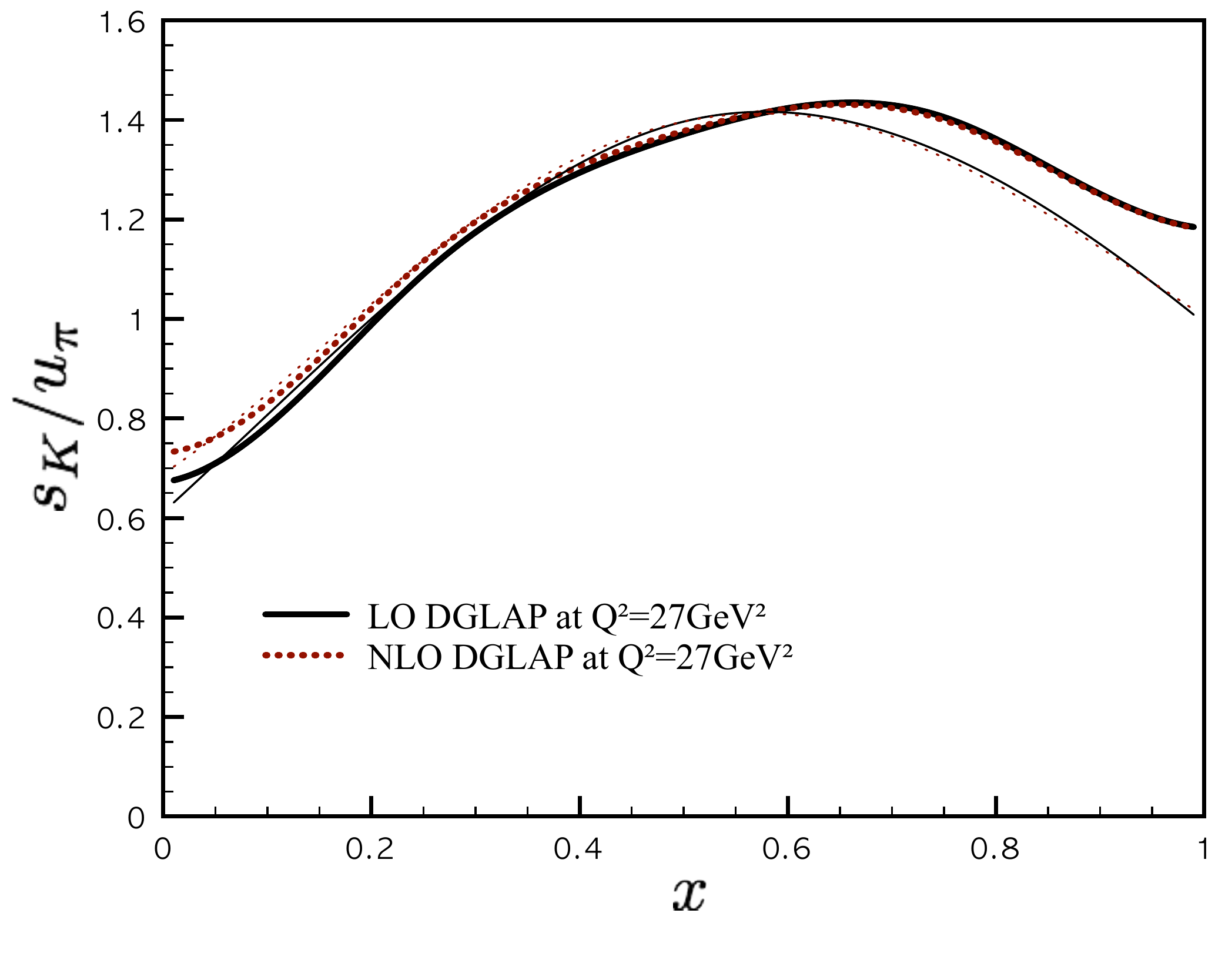}
\end{tabular}
\caption{(Color online) Ratios of the valance-quark distributions, $u_K/u_\pi$ (left) and $s_K/u_\pi$ (right) via LO and NLO DGLAP evolutions
at $Q^2=27\,\mathrm{GeV}^2$. In the right panel, we show the experimental data from Ref.~\cite{Badier:1980jq} and the fitted curve (dash)
from Ref.~\cite{Holt:2010vj}. The thick and thin lines indicate the results from Model I and II, respectively.}
\label{FIG1415}
\end{figure}

For simplicity we introduce a notation for the valance-quark distribution function as follows:
\begin{equation}
\label{eq:}
q_\phi(x)\equiv x\left[f^\phi_q(x)-f^\phi_{\bar{q}}(x) \right],
\end{equation}
where $q=(u,d,s)$ and $\phi=(\pi,K)$. In Fig.~\ref{FIG1415}, we show the numerical results for $u_K/u_\pi$ (left) and
$s_K/u_\pi$ (right) as functions of $x$. The results from Model I and II are depicted in the thick and thin lines, respectively. The results
are evolved to $Q^2=27\,\mathrm{GeV}^2$ by the LO (solid) and NLO (dotted) DGLAP evolutions. The experimental data given in the left panel of
Fig.~\ref{FIG1415} are taken from the $150\,\mathrm{GeV}$ incident-beam experiment for $(K^-,\pi^-)+\mathrm{nucleus}\to\mu^+\mu^-X$~\cite{Badier:1980jq}.
A fitted curve is also given (dashed): $u_K/u_\pi=1.1(1-x)^{0.22}$~\cite{Holt:2010vj}.

As for $u_K/u_\pi$, the numerical result reproduces the data qualitatively well.
The curves decrease with respect to $x$ when $x\gtrsim 0.4$. The results of Model I show considerable underestimate for $x\gtrsim0.5$. As for the smaller $m_s$ in Model II, the numerical results are enhanced and become close to the data points. We also note that the LO and NLO results have negligible difference. It was argued that the value of $u_K/u_\pi$ around $x=1$ manifests the nonperturbative nature of the distribution function, since it is not affected by the high-$Q^2$ evolution~\cite{Shigetani:1993dx,Holt:2010vj,Nguyen:2011jy}. In the literatures, the value is given as $u_K(x\rightarrow 1)/u_\pi(x\rightarrow 1)=(0.2\sim0.5)$. From our numerical results for Model I, we have $u_K(1)/u_\pi(1)\approx0.25$, whereas it becomes for $0.31$ Model II. Analytically, the quark-distribution function in the present theoretical framework at $x=1$ reads from Eq.~(\ref{eq:}):
\begin{equation}
\label{eq:PDF00}
f^{\phi}_{q}(x\rightarrow 1,\mu)\propto\int^\infty_0
\frac{\mathcal{M}_k\mathcal{M}_{r}}{2F^2_\phi}
\frac{\bm{k}_T\,d\bm{k}_T}
{\bm{k}^2_T+M^2_0}\left[1-\frac{2m_{q}M_0}{\bm{k}^2_T+M^2_0} \right],
\end{equation}
where $q=(u,s)$ for $\phi=(\pi,K)$. Using the above equation and the values for Model I in Eqs.~(\ref{eq:WDC1}) and (\ref{eq:WDC2}),
we can make a rough estimation for the ratio as follows:
\begin{equation}
\label{eq:RATIO}
\frac{u_K(x\rightarrow 1)}{u_\pi(x\rightarrow 1)}\approx\frac{F^2_\pi}{F^2_K}\left[1-\frac{(m_s-m_u) }{M_0}\right]=0.32,
\end{equation}
which is compatible with $0.25$. Moreover, it is clear that the ratio in Eq.~(\ref{eq:RATIO}) is a function of typical nonperturbative
quantities in QCD, such as the PS-meson weak-decay constants $F_\phi$ and constituent-quark mass $M_0$, in addition to the flavor SU(3) symmetry-breaking effects $m_u\ll m_s$. It is worth mentioning that different forms for the ratio were suggested as $(F_\pi/F_K)(M_u/M_s)^4$
from the BSE calculation~\cite{Nguyen:2011jy} and $(M_u/M_s)^2$ from the NJL calculation~\cite{Shigetani:1993dx}. In the right panel of Fig.~\ref{FIG1415},
the ratio $s_K/u_\pi$ is given as a function of $x$. The behavior of the curves are very different from those for $u_K/u_\pi$ which can be
understood easily from Fig.~\ref{FIG1213}. The value at $x=1$ becomes $1.18$ and $1.01$ for Model I and II, respectively, which is about $(3\sim5)$
times larger than that for $u_K/u_\pi$. This larger ratio is caused by that the second term in the square bracket in Eq.~(\ref{eq:RATIO})
goes to zero for $s_K/u_\pi$, i.e. $q=u$ for the both of the valance-quark distribution functions. Approximately,
using Eqs.~(\ref{eq:RATIO}), (\ref{eq:WDC1}), and (\ref{eq:WDC2}) for Model I, we obtain
$s_K(x\rightarrow 1)/u_\pi(x\rightarrow 1)\approx 0.84$, which is still compatible with the numerical result.
\section{Summary and outlook}
We have investigated the fragmentation functions of the pion and the kaon employing NLChQM in the light-cone coordinate and
apply the DLY relation to extract the quark-distribution functions from the fragmentation functions.
Two patterns of SU(3) flavor breaking have been considered: one is for the current-quark mass $(m_u,m_d,m_s)=(5,5,150)$ MeV (Model I)
and another is for $(m_u,m_d,m_s)=(2.5,5,100)$ MeV (Model II). In what follows, we list the important procedures and observations in the present work:
\begin{itemize}
\item $F_{\pi,K}$, are determined by the normalization conditions for the valence-quark distribution functions.
By doing so, we compensate the absence of the nonlocal contributions for the vector-current conservation phenomenologically.
The calculated values of $F_{\pi,K}$ turn out to be about $(5\sim10)\%$ smaller from their empirical values.
\item The curves of the quark-distribution function, multiplied by $x$, i.e. $xf^\phi_q(x)$ for the pion and kaon exhibit very different shapes.
It is due to the explicit flavor SU(3) symmetry breaking. Moreover, the shapes of those curves are substantially different from the ones generated from usual local-interaction models.
\item The curves of the elementary fragmentation functions of the pion and kaon, multiplied by $z$, i.e. $zd^{\phi}_{q}$ are also different.
The pion one is symmetric around $z=1/2$ but the kaon ones are titled to higher $z$ side. Furthermore the magnitude of the pion one is large more larger than the kaon ones.
\item For the renormalized fragmentation functions of the pion and kaon, multiplied by $z$, i.e. $zD^{\phi}_{q}$, we find $zD^{K}_{\bar{s}}$ are much larger than $zD^{K}_{u}$. This is quite different from the results from other models with the local couplings.
\item The kaon fragmentation functions are evolved to $Q^2=1\,\mathrm{GeV}^2$. Compared with the empirical curves our curves are much larger.
      We also notice that $zD^{K}_{u}$ is more sensitive to the strange quark mass than $zD^{K}_{\bar{s}}$. 
\item The plus-type and minus-type quark-distribution functions are evolved to $Q^2=4\,\mathrm{GeV}^2$.
They are compared with the empirical data for the pion, resulting in a qualitative good agreement with them.
We also computed those for the kaon within the same framework. The fragmentation functions for the kaon are evolved to $Q^2=1\,\mathrm{GeV}^2$ and we observed that strong overshoots in the vicinity of $z=0.5$ compared with the empirical data.
\item We presented the valance-quark distribution functions at $Q^2=27\,\mathrm{GeV}^2$ for the pion and kaon.
The numerical results are qualitatively compatible with the empirical data for the pion. We also calculated the ratios for
the valance-quark distributions for the kaon and pion. Although the numerical results showed qualitative agreement with the data,
there appears considerable deviations for the region $x\gtrsim0.5$ for $u_K/u_\pi$. Theoretical estimations for $s_K/u_\pi$ are presented.
\end{itemize}

As noticed in Section III, we have not taken into account the axial-current conservation in the present framework, which may
become problematic for the nonlocal quark-PS meson interactions as in NLChQM~\cite{Nam:2006sx}. Moreover, we also note that, in the present
framework, the momentum-dependencies in the effective quark masses in the fragmentation and distribution functions, given in Eqs.~(\ref{eq:DDDDD})
and (\ref{eq:PDF0}), were simplified to avoid complexities in numerical calculations, i.e. $M_\ell\to M_0$. In order for theoretical consistencies,
these two issues might be taken into account carefully, although the present results are phenomenologically acceptable by comparison with the data.
We verified that the (valance) quark-distribution functions are considerably modified by correcting those issues, satisfying relevant theoretical constraints, and the same for the fragmentation functions accordingly~\cite{NAMKAO}. Related works are under progress and appear elsewhere.
\section*{Acknowledgments}
S.i.N. is very grateful to the hospitality during his visiting National Taiwan University (NTU) with the financial support from NCTS (North) in Taiwan, where the present work was performed. The work of C.W.K. was supported by the grant NSC 99-2112-M-033-004-MY3 from National Science Council (NSC) of Taiwan. He has also acknowledged the support of NCTS (North) of Taiwan.
\section*{Appendix}
\begin{table}[h]
\begin{tabular}{c|ccccccc}
$\mathcal{C}^\phi_q$&$\pi^0$&$\pi^+$&$\pi^-$&$K^0$&$\bar{K}^0$&$K^+$&$K^-$\\
\hline
$u$&$1/2$&$1$&$0$&$0$&$0$&$1$&$0$\\
$d$&$1/2$&$0$&$1$&$1$&$0$&$0$&$0$\\
$s$&$0$&$0$&$0$&$0$&$1$&$0$&$1$\\
$\bar{u}$&$1/2$&$0$&$1$&$0$&$0$&$0$&$1$\\
$\bar{d}$&$1/2$&$1$&$0$&$0$&$1$&$0$&$0$\\
$\bar{s}$&$0$&$0$&$0$&$1$&$0$&$1$&$0$\\
\end{tabular}
\caption{Flavor factors in Eq.~(\ref{eq:FRAG}).}
\label{TABLE0}
\end{table}


\begin{thebibliography}{99}
\bibitem{Collins:1992kk}
  J.~C.~Collins,
  Nucl.\ Phys.\  {\bf B396}, 161 (1993).
\bibitem{Mulders:1995dh}
  P.~J.~Mulders, R.~D.~Tangerman,
  Nucl.\ Phys.\  {\bf B461}, 197 (1996).
\bibitem{Boer:1997nt}
  D.~Boer, P.~J.~Mulders,
  Phys.\ Rev.\  {\bf D57}, 5780 (1998).
\bibitem{Anselmino:1994tv}
  M.~Anselmino, M.~Boglione, F.~Murgia,
  Phys.\ Lett.\  {\bf B362}, 164 (1995).
\bibitem{Anselmino:2008jk}
  M.~Anselmino {\it et al.},
  Nucl.\ Phys.\ Proc.\ Suppl.\  {\bf 191}, 98 (2009).
\bibitem{Christova:2006qs}
  E.~Christova and E.~Leader,
  Eur.\ Phys.\ J.\ C {\bf 51}, 825 (2007).
\bibitem{Anselmino:2007fs}
  M.~Anselmino {\it et al.},
  Phys.\ Rev.\  D {\bf 75}, 054032 (2007).
\bibitem{Bacchetta:2006tn}
  A.~Bacchetta, M.~Diehl, K.~Goeke, A.~Metz, P.~J.~Mulders and M.~Schlegel,
  JHEP {\bf 0702}, 093 (2007).
\bibitem{Efremov:2006qm}
  A.~V.~Efremov, K.~Goeke and P.~Schweitzer,
  Phys.\ Rev.\  D {\bf 73}, 094025 (2006).
\bibitem{Collins:2005ie}
  J.~C.~Collins, A.~V.~Efremov, K.~Goeke, S.~Menzel, A.~Metz and P.~Schweitzer,
  Phys.\ Rev.\  D {\bf 73}, 014021 (2006).
\bibitem{Ji:2004wu}
  X.~d.~Ji, J.~p.~Ma and F.~Yuan,
  Phys.\ Rev.\  D {\bf 71}, 034005 (2005).
\bibitem{Drell:1969jm}
  S.~D.~Drell, D.~J.~Levy, T.~-M.~Yan,
  Phys.\ Rev.\  {\bf 187}, 2159 (1969).
\bibitem{Kretzer:2000yf}
  S.~Kretzer,
  Phys.\ Rev.\  {\bf D62}, 054001 (2000).
\bibitem{Hirai:2007cx}
  M.~Hirai, S.~Kumano, T.~H.~Nagai and K.~Sudoh,
  Phys.\ Rev.\  D {\bf 75}, 094009 (2007).
\bibitem{Kniehl:2000fe}
  B.~A.~Kniehl, G.~Kramer, B.~Potter,
  Nucl.\ Phys.\  {\bf B582}, 514 (2000).
\bibitem{Conway:1989fs}
  J.~S.~Conway {et al.},
  Phys.\ Rev.\  {\bf D39}, 92 (1989).
\bibitem{deFlorian:2007aj}
  D.~de Florian, R.~Sassot and M.~Stratmann,
  Phys.\ Rev.\ D {\bf 75}, 114010 (2007).
\bibitem{Sutton:1991ay}
  P.~J.~Sutton, A.~D.~Martin, R.~G.~Roberts, W.~J.~Stirling,
  Phys.\ Rev.\  {\bf D45}, 2349 (1992).
 \bibitem{Ji:1993qx}
  X.~-D.~Ji and Z.~-K.~Zhu,
  [hep-ph/9402303].
\bibitem{Jakob:1997wg}
  R.~Jakob, P.~J.~Mulders and J.~Rodrigues,
  Nucl.\ Phys.\ A {\bf 626}, 937 (1997).
\bibitem{Meissner:2010cc}
  S.~Meissner, A.~Metz and D.~Pitonyak,
  Phys.\ Lett.\  B {\bf 690}, 296 (2010).
\bibitem{Bentz:1999gx}
  W.~Bentz, T.~Hama, T.~Matsuki, K.~Yazaki,
  Nucl.\ Phys.\  {\bf A651}, 143 (1999).
\bibitem{Ito:2009zc}
  T.~Ito, W.~Bentz, I.~-Ch.~Cloet, A.~W.~Thomas and K.~Yazaki,
  Phys.\ Rev.\  D {\bf 80}, 074008 (2009).
\bibitem{Matevosyan:2010hh}
  H.~H.~Matevosyan, A.~W.~Thomas and W.~Bentz,
  Phys.\ Rev.\  D {\bf 83}, 074003 (2011)
\bibitem{Matevosyan:2011ey}
  H.~H.~Matevosyan, A.~W.~Thomas and W.~Bentz,
  Phys.\ Rev.\  D {\bf 83}, 114010 (2011).
\bibitem{Matevosyan:2011vj}
  H.~H.~Matevosyan, W.~Bentz, I.~C.~Cloet and A.~W.~Thomas,
  Phys.\ Rev.\  D {\bf 85}, 014021 (2012).
\bibitem{Nguyen:2011jy}
  T.~Nguyen, A.~Bashir, C.~D.~Roberts and P.~C.~Tandy,
  Phys.\ Rev.\ C {\bf 83}, 062201 (2011).
\bibitem{Noguera:2005cc}
  S.~Noguera and V.~Vento,
  Eur.\ Phys.\ J.\ A {\bf 28}, 227 (2006).
\bibitem{Aicher:2010cb}
  M.~Aicher, A.~Schafer and W.~Vogelsang,
  Phys.\ Rev.\ Lett.\  {\bf 105}, 252003 (2010).
\bibitem{Holt:2010vj}
  R.~J.~Holt and C.~D.~Roberts,
  Rev.\ Mod.\ Phys.\  {\bf 82}, 2991 (2010).
\bibitem{Dorokhov:1991nj}
  A.~E.~Dorokhov,
  Nuovo Cim.\  A {\bf 109}, 391 (1996).
\bibitem{Praszalowicz:2001pi}
  M.~Praszalowicz and A.~Rostworowski,
  Phys.\ Rev.\  D {\bf 66}, 054002 (2002).
\bibitem{Nam:2006sx}
  S.~i.~Nam and H.~-Ch.~Kim,
  Phys.\ Rev.\  D {\bf 74}, 076005 (2006).
\bibitem{Nam:2006au}
  S.~i.~Nam, H.~-Ch.~Kim, A.~Hosaka and M.~M.~Musakhanov,
  Phys.\ Rev.\  D {\bf 74}, 014019 (2006).
\bibitem{Bacchetta:2002tk}
  A.~Bacchetta, R.~Kundu, A.~Metz and P.~J.~Mulders,
  Phys.\ Rev.\  D {\bf 65}, 094021 (2002).
\bibitem{Amrath:2005gv}
  D.~Amrath, A.~Bacchetta and A.~Metz,
  Phys.\ Rev.\  D {\bf 71}, 114018 (2005).
\bibitem{Bacchetta:2007wc}
  A.~Bacchetta, L.~P.~Gamberg, G.~R.~Goldstein and A.~Mukherjee,
  Phys.\ Lett.\ B {\bf 659}, 234 (2008).
\bibitem{Bacchetta:2006un}
  A.~Bacchetta and M.~Radici,
  Phys.\ Rev.\  D {\bf 74}, 114007 (2006).
\bibitem{Aloisio:2003xj}
  R.~Aloisio, V.~Berezinsky, M.~Kachelriess,
  Phys.\ Rev.\  {\bf D69}, 094023 (2004).
\bibitem{Nam:2011hg}
  S.~i.~Nam and C.~W.~Kao,
  Phys.\ Rev.\  D {\bf 85}, 034023 (2012).
\bibitem{Shuryak:1981ff}
E.~V.~Shuryak,
Nucl.\ Phys.\ B \textbf{203}, 93 (1982).
\bibitem{Diakonov:1985eg}
  D.~Diakonov and V.~Y.~Petrov,
  Nucl.\ Phys.\  B \textbf{272}, 457 (1986).
\bibitem{Diakonov:1983hh}
D.~Diakonov and V.~Y.~Petrov,
Nucl.\ Phys.\ B \textbf{245}, 259 (1984).
\bibitem{Schafer:1996wv}
T.~Sch\"afer and E.~V.~Shuryak,
Rev.\ Mod.\ Phys.\ \textbf{70}, 323 (1998).
\bibitem{Diakonov:2002fq}
D.~Diakonov,
Prog.\ Part.\ Nucl.\ Phys.\ \textbf{51}, 173 (2003).
\bibitem{Musakhanov:1998wp}
  M.~Musakhanov,
  Eur.\ Phys.\ J.\  {\bf C9 }, 235 (1999).
\bibitem{Musakhanov:2002vu}
  M.~Musakhanov,
  Nucl.\ Phys.\ A {\bf 699}, 340 (2002).
\bibitem{Nam:2007gf}
  S.~i.~Nam and H.~-Ch.~Kim,
  Phys.\ Rev.\  D \textbf{77}, 094014 (2008).
\bibitem{Nam:2010pt}
  S.~i.~Nam and H.~-Ch.~Kim,
  Phys.\ Lett.\  B {\bf 700}, 305  (2011).
\bibitem{Dorokhov:2000gu}
  A.~E.~Dorokhov, L.~Tomio,
  Phys.\ Rev.\  {\bf D62}, 014016 (2000).
\bibitem{Dorokhov:2002iu}
  A.~E.~Dorokhov,
  JETP Lett.\  {\bf 77}, 63 (2003).
\bibitem{Botje:2010ay}
  M.~Botje,
  Comput.\ Phys.\ Commun.\  {\bf 182}, 490 (2011).
\bibitem{DGLAP}
The QCD-evolution fortran program QCDNUM17, written by M.~Botje, at http://www.nikhef.nl/user/h24/qcdnum.
\bibitem{Aybat:2011zv}
  S.~M.~Aybat and T.~C.~Rogers,
  Phys.\ Rev.\ D {\bf 83}, 114042 (2011).
\bibitem{Shigetani:1993dx}
  T.~Shigetani, K.~Suzuki and H.~Toki,
  Phys.\ Lett.\  B {\bf 308} 383, (1993).
\bibitem{Nakamura:2010zzi}
  K.~Nakamura {\it et al.} [Particle Data Group Collaboration],
  J.\ Phys.\ G {\bf 37}, 075021 (2010).
\bibitem{Nam:2008bq}
  S.~i.~Nam,
  Phys.\ Rev.\  D {\bf 79}, 014008 (2009).
\bibitem{Badier:1980jq}
  J.~Badier {\it et al.}
  [Saclay-CERN-College de France-Ecole Poly-Orsay Collaboration],
  Phys.\ Lett.\  B {\bf 93}, 354 (1980).
\bibitem{NAMKAO}
  S.~i.~Nam and C.~W.~Kao, in preparation.
\end{thebibliography}
\end{document}